# A Review of the Degradation Mechanisms of NCM Cathodes and Corresponding Mitigation Strategies

Liga Britala[1,2], Mario Marinaro[3], Gints Kucinskis[1]*

[1]Institute of Solid State Physics, University of Latvia, 8 Kengaraga Street, Riga, LV-1063, Latvia

[2]Faculty of Materials Science and Applied Chemistry, Riga Technical University, 3/7 Paula Valdena Street, Riga, LV-1048, Latvia

[3]ZSW-Zentrum für Sonnenenergie und Wasserstoff-Forschung, Helmholtzstrasse 8, Ulm, 89081, Germany

*e-mail of the corresponding author: gints.kucinskis@cfi.lu.lv

## Abstract

Li-ion batteries (LIBs) are the most widely used form of energy storage in mobile electronic devices and electric vehicles. Li-ion battery cathodes with the composition $LiNi_xMn_yCo_zO_2$ (NCMs) currently display some of the most promising electrochemical characteristics for high performance LIBs. NCM compositions with high nickel content (x>0.8) exhibit the largest specific capacity while undergoing fast degradation and presenting safety issues. As the main degradation mechanisms of NCM materials and the mitigation of their degradation, are still subjects of many ongoing studies, this work summarizes the current knowledge on the subject. Here, the existing literature is reviewed to present the structural and electrochemical degradation of NCM with varying Ni stoichiometries (NCM111, NCM622, NCM811, and beyond). Routes for hindering the degradation of NCM are discussed as a function of Ni content in NCM and include doping, application of protective coatings, and engineering of the microstructure. A comprehensive understanding of the main degradation pathways of NCM is key to applying the most appropriate mitigation strategies and keep advancing towards higher energy NCM materials with longer cycle-life.

# Contents

# 1. INTRODUCTION

Li-ion batteries (LIBs) have been at the forefront of mobile energy storage since their commercialization by Sony in 1991 due to their high capacity, excellent energy density, and good safety. The capacity and hence energy of a LIB is largely determined by the cathode of the battery, since, on average, it occupies roughly 41 % of the total weight of the battery cell [1]. Layered transition metal oxide materials with the general formula $LiNi_xMn_yCo_zO_2$ (NCM, x+y+z=1), first reported in 1999[2], have become one of the most studied and commercially used cathode materials for Li-ion batteries[3]. NCM is an evolution of layered $LiCoO_2$ first reported in 1980 and has gained popularity due to its high capacity, energy density, and good stability. Gradual increases in the content of Ni have enabled further increases in the specific capacity of NCM materials while alleviating some of the sustainability and cost concerns due to the reduced cobalt content. Currently, Ni-rich cathodes with the composition $LiNi_{0.8}Mn_{0.1}Co_{0.1}O_2$ (NCM 811) have become increasingly popular and are being incorporated in many of the state-of-art batteries for electric vehicles (EVs) and portable electronics[4]. NCMs with higher Ni content are under development and, provided that some material stability-related challenges are solved, are expected to appear more prominently in the market over the coming years.

Increasing Ni content in NCM has clear benefits regarding improved specific energy and energy density. However, the safety and cycle life of NCMs with higher Ni content can suffer dramatically, putting further increases in energies at odds with the cycle life. Suggested decomposition pathways for the ageing of NCM range from microcracking and undesired phase transitions to transition metal (TM) dissolution, which in turn causes anodic solid-electrolyte interphase (SEI) degradation and loss of lithium inventory (LLI)[5–9]. The decomposition pathways may vary as the nickel content in NCM ranges from x=0.33 up to x=0.95, further complicating the generalization of the current knowledge. Thus, the dominant pathways and precise mechanisms for capacity fading of NCM cathodes remain elusive. Yet, this knowledge is crucial for the continued development of NCM materials with high energy density and long cycle life.

To counter the degradation of NCM cells and make the NCM active materials more stable, common strategies include tailoring the electrode formulation and grain microstructure, modifying the crystal lattice by introducing dopants or coating the active material. The latter often forms an artificial layer of cathode-electrolyte interphase (CEI) that prevents undesired reactions between cathode and electrolyte[10,11]. At the same time, doping mitigates the lattice expansion and can have further effects on the conductivity and structural stability of NCM. However, despite significant progress in improving the cycle life of NCM-based batteries[12], further improvements are still required. To effectively counter the limitations of Ni-rich cathodes for further commercialization, it is important first to fully understand the degradation mechanisms of these materials. This calls for a more structured and evidence-based approach in searching for and optimizing the cycle life of NCM materials. Yet most of the existing review articles focus on synthesis[13–16], structural modification, doping, and coating[17–20], ageing[21–24], or other aspects[25–31] exclusively but not in sufficient context of each other. Moreover, although some high-quality reviews have been made available before[23,32–36], many fundamental studies released over the last years have substantially altered the view on redox activity, ageing, and decomposition of NCM cathodes[37–40], thus justifying an updated summary of the topic. Furthermore, enough experimental studies have appeared to enable a statistical analysis of the practical benefits of structural modification methods used to mitigate the ageing.

In this review, we thoroughly and systematically analyse existing scientific studies to unravel the main degradation pathways of NCM cathodes based on the Ni content. Strategies for mitigating ageing as a function of Ni content in NCM are then surveyed and reviewed, basing the conclusions on a statistical analysis of existing empirical studies and relating the mitigation strategies to the dominant ageing mechanisms.

## 2. DEGRADATION MECHANISMS IN NCM CATHODES

### 2.1. Overview of Degradation Mechanisms

NCM cathode active materials (CAMs) have the general chemical composition of $LiNi_xMn_yCo_zO_2$. Analogously to $LiCoO_2$[41], NCMs have an O3 layered structure (O3 signifies lithium in the octahedral site and 3 lithium and transition metal – oxygen layers in a repeat unit[42]) and belong to the $R3m$ space group[43]. The primary particles of these cathode materials consist of alternating TM and Li layers within which atoms are arranged hexagonally and are enclosed by face-sharing oxygen octahedra (figure 1a). Lithium ions can freely diffuse along the 2D Li plane from one octahedral site to another via adjacent tetrahedral sites[44].

$LiCoO_2$ (LCO) has historically been the first oxide cathode developed for Li-ion batteries[45]. It is likely the most widely used cathode active material since the first successful commercialization of Li-ion batteries in 1991[46]. However, delithiating LCO by more than 50 % leads to the release of oxygen, as the $Co^{3+/4+}$ band overlaps with the top of the $O^{2-}$:2p band, causing irreversible oxygen oxidation and structural collapse at high degrees of delithiation[47]. This limits the practically obtainable capacity to a modest ≈140 mAh/g and has promoted a search for compositions that substitute Co with other metals. Among these, NCM materials have gained popularity. While $LiMnO_2$ is impossible to synthesize in the desired O3 layered structure and $LiNiO_2$ suffers from structural and chemical stability issues, the combination of Ni, Co, and Mn leads to improved energy density and structural stability. The use of three transition metal ions ensures Ni redox reactions at potentials below those of $O^{2-}$:2p band while Mn serves as a structural stabilizer and Co further improves structural stability and electrical conductivity[47]. Price considerations, political issues, and low abundance of Co, along with increasing demand for higher energy density, drive the composition of NCM towards Ni-rich materials. Hence, NCM materials have gradually developed from NCM111, with the specific composition first reported in 2001 by T. Ohzuku et al.[48], to NCM532, NCM622, NCM811, and beyond (numbers denote ratio between Ni, Co, and Mn). Unfortunately, while the energy density increases with growing Ni content, high-Ni NCMs also suffer from a comparatively more rapid capacity fade during cycling (figure 1b). This drives the efforts to understand and extend the rapidly diminishing cycle life of Ni-rich NCM while exploiting the attractively high energy density that these materials provide.

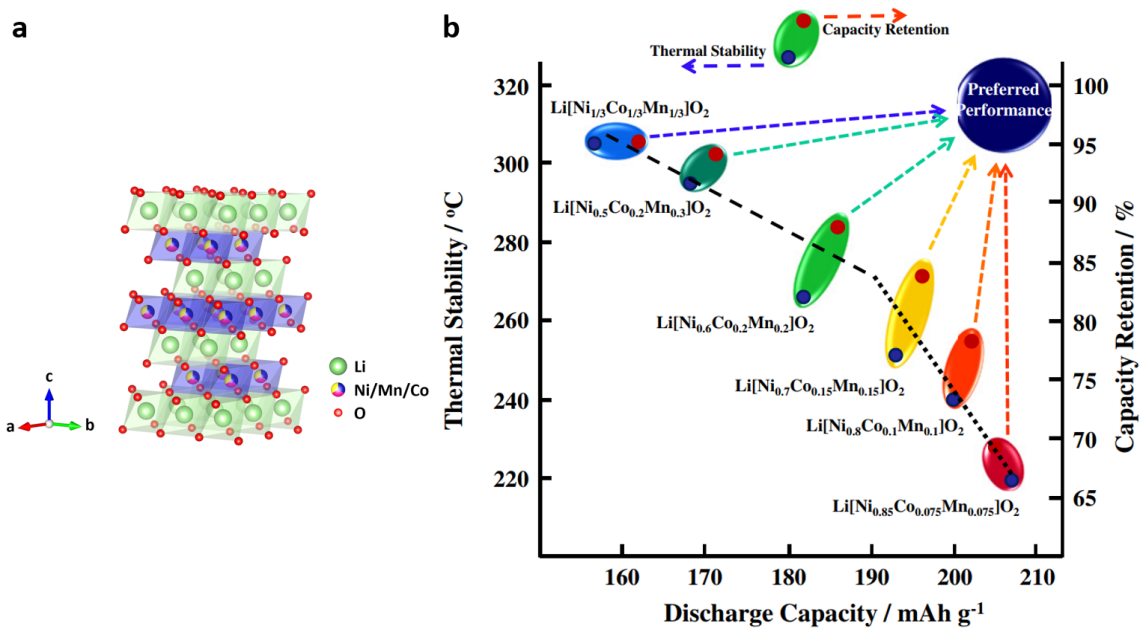

**Figure 1.** a) Crystallographic structure of NCM, b) Discharge capacity, capacity retention, and thermal stability of NCMs with different Ni composition. Reprinted with permission from H.-J. Noh et al.[49], copyright 2013 Elsevier.

The degradation mechanisms related to the NCM cathode are outlined in figure 2 and mainly stem from two distinct sources – the synthesis of cathode material and the operation of battery (lithiation-delithiation of cathode) once the cathode is incorporated within the cell. Both are reviewed in detail in the subsequent chapters.

Firstly, during the synthesis and handling of the material before the assembly of a battery cell, attention should be paid to controlling lithium content preventing the formation of residual lithium compounds (RLCs) on the surface of the NCM material. These can lead to a loss of lithium inventory in the assembled cell and undesired reactions between RLCs and electrolyte. Secondly, and perhaps most importantly, NCM undergoes degradation during cycling - there, the degradation mechanisms are arguably more complex.

Among the main degradation mechanisms for any NCM cathode material during cycling are cathode-electrolyte interphase (CEI) formation, surface reconstruction, oxygen release, TM dissolution, and microcracking, forming new reactive surfaces. Most degradation mechanisms are not occurring individually. They are profoundly interconnected and often trigger one another. The multitude of decomposition mechanisms (some triggering secondary degradation responses) and the abundance of various NCM compositions are the main reasons why the relative contribution and importance of each degradation mechanism has thus far remained relatively little understood in terms of quantifiable parameters.

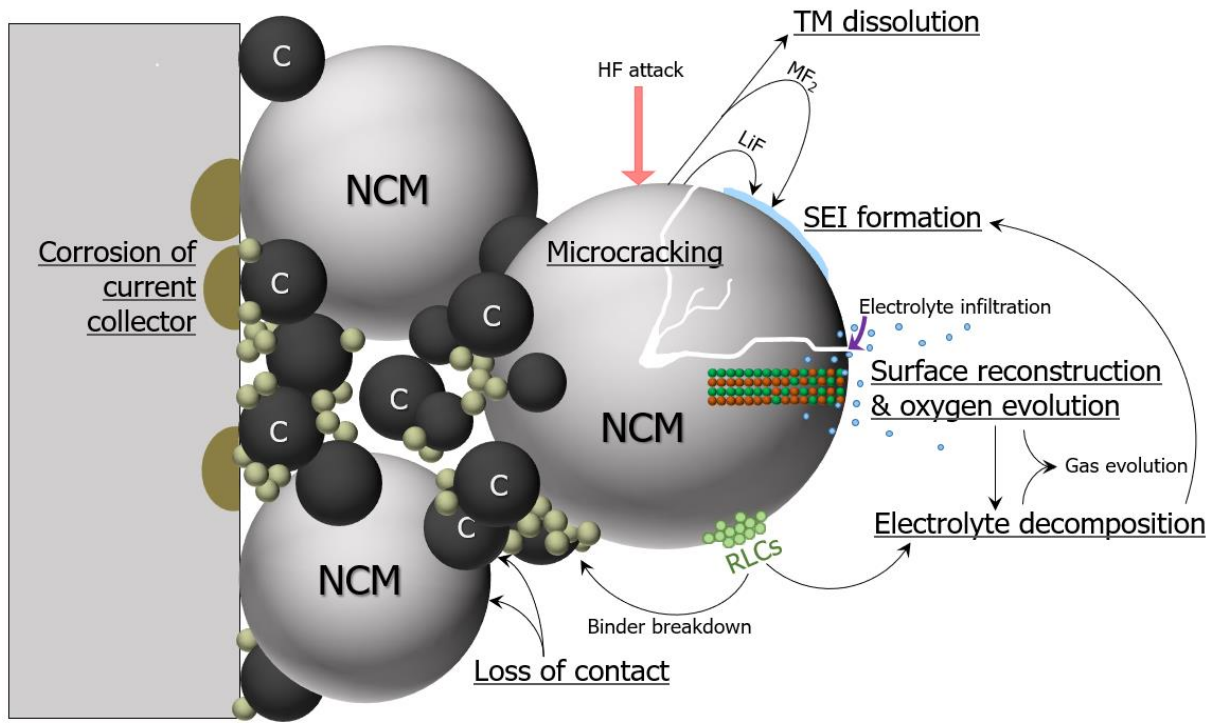

**Figure 2**. Ageing mechanisms of NCM cathode materials.

## 2.2. Degradation mechanisms from synthesis and storage

Several different methods have been developed for the synthesis of NCM, including sol-gel, hydrothermal, solid-state, and pulse combustion[50]. However, the most often employed method is co-precipitation during which an NCM precursor is formed by co-precipitation of stoichiometric amounts of TM salts (usually sulphates) chelated by ammonia and reacted with hydroxide ions in a water solution[15]. The obtained, dried TM hydroxide precursor is then mechanically mixed with a lithium source (usually LiOH·$H_2$O or Li$CO_3$) and then sintered at temperatures around 900°C to obtain the final LiNi$_x$Co$_y$Mn$_z$O$_2$ product.

At elevated temperatures lithium sublimation occurs due to the high vapor pressure of lithium, thus allowing the nickel ions to move to the lithium sites and increasing cation mixing[51,52]. To counter these negative effects and achieve the desired performance excess lithium is usually added upon synthesis of NCM. Furthermore, a larger amount of lithium source is customarily added to NCM with higher Ni content to replace the Ni which readily occupies Li sites and counter the Li sublimation during synthesis. As the same amount of Li atoms can be inserted into the lattice of NCM111 as NCM811, but more lithium is added during synthesis of NCM811, more lithium remains on the surface of the NCM811 particles. Correlation between the amount of nickel in NCM and RLCs on particle surface has also been reported[49,51,53].

Ni-rich NCM active materials are particularly sensitive to moisture ($H_2O$) and carbon dioxide ($CO_2$)[54], owing to the alkalinity of the material increasing with larger nickel content[55]. $Li^+/H^+$ exchange can occur due to wet processing or exposure to ambient moisture[56–59]. These factors can also contribute to the formation of RLCs, as extracted lithium resides on the surface, reacting with $O_2$, $CO_2$, and $H_2O$, forming $Li_2O$ and the more stable $Li_2CO_3$ and LiOH.

There are several side reactions that RLCs residing on the surface of NCM can partake in a battery cell. These are highly detrimental to battery efficiency:

> 1) $Li_2CO_3$ electrochemical decomposition via peroxodicarbonate intermediate to form singlet oxygen: $2Li_2CO_3 \rightarrow 4Li^+ + 4e^- + 2CO_2 + {}^1O_2$ (>3.8V)[60]

2) LiOH reaction with PVDF binder: LiOH + (CH$_2$-CF$_2$)$_n$ → (CH=CF)$_n$ + LiF + H$_2$O

3) LiOH reaction with LiPF$_6$ electrolyte: LiOH + LiPF$_6$ → 3LiF + PF$_3$O + H$_2$O

4) Li$_2$O reaction with H$_2$O formed upon electrolyte decomposition: Li$_2$O + H$_2$O → 2LiOH

5) RLC reactions with HF formed upon electrolyte decomposition:

    a) Li$_2$CO$_3$ + 2HF → CO$_2$ + H$_2$O + 2LiF

    b) LiOH + HF → LiF + H$_2$O

    c) Li$_2$O + HF → LiF + H$_2$O

HF is usually present due to the decomposition of LiPF$_6$ electrolyte salt (reactions 6-8). If LiPF$_6$ is used as the conducting salt in the electrolyte, water, if any, resides on the cathode's surface or is formed in RLC reactions, will react with the electrolyte salt and further increase the acidity and reactivity of the environment by producing HF:

6) LiPF$_6$ + H$_2$O → LiF + PF$_3$O + 2HF

7) PF$_3$O + H$_2$O → HF + HPO$_2$F$_2$

8) LiPF$_6$ + H$^+$ → Li$^+$ + PF$_5$ + HF

The HF produced from electrolyte decomposition will attack the RLCs on the cathode surface (reaction 5) and the active material (TM oxides). The reaction produces slightly more soluble TMF$_2$ salts, which can dissolve in the electrolyte solution, diffuse to the anode, and deteriorate the solid-electrolyte interphase (SEI) or remain on the cathode to participate in CEI formation[53,61]. The presence of HF can also lead to the corrosion of the copper current collector[62] as opposed to the aluminium current collector, where it reacts with the passivating native Al$_2$O$_3$ layer forming an additional protective AlF$_3$ layer that hinders further Al corrosion[62,63]. Conversely, in alkaline (pH > 11) solutions in the presence of water (water-based NCM, NCA, or LFP slurries) protective Al$_2$O$_3$ from the surface of the current collector can dissolve as Al(OH)$_4^-$ resulting in erosion of Al foil and contact loss[63].

By producing $H_2O$, reactive singlet oxygen $^1O_2$, and various gases ($CO_2$, CO, $O_2$) upon reactions with the battery environment, RLCs promote electrolyte degradation, further RLC reactions, reactions with the active material, and CEI formation, resulting in increased impedance and cell bloating, which can lead to thermal runaway and compromise battery safety. Experiments have shown that the main component of gases released at the NCM cathode side is $CO_2$, followed by CO, and, at a lower level, also $O_2$[64]. For more details on gas evolution experiments in liquid and solid-electrolyte-based batteries, the reader is referred to the review article in reference[64].

## 2.3. Degradation mechanisms from operation

The speed and type of degradation that stems from the operation of NCM cathodes, namely lithiation, and delithiation, depends on many parameters, including the voltage range and temperature in which the battery is cycled and the relative amounts of TMs in the active material (predominantly Ni content). The most significant chemical and structural changes occur at high degrees of delithiation (high voltages) and with increasing Ni content in the active material.

As the voltage increases during battery charging, different degradation phenomena come into play, based on the degree of delithiation or state of charge (SoC) of the cathode. Here, we use the term SoC, with 100% SoC signifying all lithium being extracted from the active cathode material and 0% SoC corresponding to stoichiometry $LiNi_xMn_yCo_zO_2$ (fully lithiated). From 0% SoC to around 56% SoC NCMs are structurally stable, and the charge compensation occurs via reversible reactions, thus, no major degradation and capacity fade is observed in the cathodes[38]. Beyond 56% SoC, it has been reported that irreversible charge compensation reactions start occurring in all NCMs[38]. However, experimental results of lattice parameter change, oxygen evolution, and capacity fade demonstrate that no major, irreversible degradation occurs up to around 70-80% SoC[39,40,65,66]. When 80% SoC is exceeded, significant degradation effects such as sudden lattice parameter change and oxygen release along with surface reconstruction start to accelerate the ageing of the cathode, which is evidenced by a rapid capacity fade of all NCMs when they are charged beyond their respective 80% SoC voltages (4.7 V for NCM111, 4.6-4.7 V for NCM622, and 4.3 V for NCM811)[37,39,40,65–67]. For cathodes with varying Ni content the voltage threshold corresponding to 80% SoC differs. For NCM811 80% SoC occurs at around 4.3 V, whereas NCM111 reaches the 80% SoC only at 4.7 V vs. Li/Li$^+$[37,40]. While the higher cell voltage might be somewhat desired due to increased energy density, 4.7 V is beyond the electrochemical stability window of most conventional liquid electrolytes. Thus, lowering the cell voltage with increasing nickel content while accessing the same amount of capacity has historically been one of the reasons why the research community and industry have gradually shifted towards NCMs with higher Ni content.

The NCM cathode degradation mechanisms can be divided into i) lattice contraction/expansion during operation which causes volume changes and microcracking, ii) phase transitions along with oxygen release, iii) TM dissolution, which results in loss of active material (LAM) and loss of lithium inventory (LLI), and iv) a passivating surface layer (CEI) growth, which increases impedance and causes LLI and LAM. The degradation occurring during battery operation is more detrimental to the cathode material stability than the RLC effect due to it affecting the cathode material beyond the first few cycles and having a more significant effect on the structural stability of the cathode. The negative impact of RLCs is notable only in the first few cycles, during which they disintegrate, get trapped within the CEI, or participate in electrolyte and binder breakdown (reactions 1-5).

The degradation mechanisms within NCMs occurring from battery operation are all directly or indirectly linked. For example, oxygen evolution along with surface reconstruction contributes to electrolyte degradation, facilitating TM dissolution and promoting CEI growth, leading to increased impedance, lowered reversible capacity, safety issues, and decreased cycle life.

### Lattice volume change and microcrack formation

NCM cathodes most often consist of secondary particles composed of agglomerated, randomly oriented primary layered particles (figure 3a). Microcracks occurring within the cathode material can be either intragranular or intergranular. Intragranular cracks (within the grain) may occur on the interfaces between the layered phase of primary particle bulk and the spinel and rock salt phases of the surface regions due to volume change in the layered phase and little to no volume change in the rigid rock salt surface phases. Intergranular cracks (cracks between primary particles) are deemed more degrading to battery performance than intragranular cracks[68].

When the cathode is delithiated, the primary particle crystal lattice expands and contracts unevenly in *a* and *c* directions with varying magnitudes based on Ni content and voltage (figure 3b,c,d)[69]. In NCM811, at voltages over 4.0 V, significant collapse of the *c* lattice parameter sets in. At 4.3 V, the *c* parameter reaches values lower than in the discharged state and keeps shrinking if discharged further. On the other hand, NCM111 shows minimal shrinkage in the *c* parameter within this voltage range (figure 3b).

However, when changes in the *c* lattice parameter are correlated with the state of charge (and not voltage), all NCM materials show a similar trend (figure 3d)[39]. For instance, at 4.3 V, NCM811 has already reached 80% SoC while NCM111 is delithiated to only 60% SoC. When charged to 80% SoC or 4.7 V, significantly above the electrochemical stability window of most liquid electrolytes, NCM111 experiences a similar collapse of the *c* lattice parameter (figure 3d). The characteristic that makes NCMs with higher Ni content appear more structurally unstable is the lower voltage at which a critical SoC is reached (80% SoC at 4.3 V for NCM811 vs. 80% SoC at 4.7 V for NCM111). Thus, from a fundamental point of view, it is perhaps not entirely informative to compare the two materials within the same voltage range when their state of lithiation and thus the degradation susceptibility is dramatically different.

The sharp change in lattice parameters generates mechanical stress on the interfaces of the randomly oriented primary particles, eventually causing them to separate, thus disintegrating the secondary particles. While this results in a larger surface area lowering the impedance and improving the rate capability of the cathode, the additional surface area also accelerates the undesired reactions with the electrolyte, surface phase transitions, and CEI growth.

Intergranular microcrack mitigation has been attempted using single crystalline NCM (materials made of single crystallites as opposed to spherical aggregates of primary particles)[70]. Although the lack of active surfaces compared to polycrystalline, cracked particles results in relatively inferior rate capability and reversible capacity, the maintained integrity of single crystalline particles is somewhat desirable, as the advantages of secondary particle cracking do not outweigh the drawbacks of rapid capacity fade and safety issues associated with it. Thus, a solution to microcracking in polycrystalline NCMs is warranted to produce a safer, cheaper, and more stable cathode material with good rate capability, reversible capacity, and cycling stability.

Avoiding the change in volume or lattice parameters is another strategy. This is a relatively challenging task and is usually reduced to the development of zero volume change (limited change in unit cell volume, although individual lattice parameters might change) or even zero strain (limited changes of lattice parameters) materials[71]. The former has been reported for NCM layered cathode material by R. Zhang et al.[72], while the latter has up to now been observed along with zero volume change only for spinel or rock-salt materials. Zero volume change and zero strain materials are typically obtained by doping – substituting some atoms with a different species (commonly transition metals replaced by other metals).

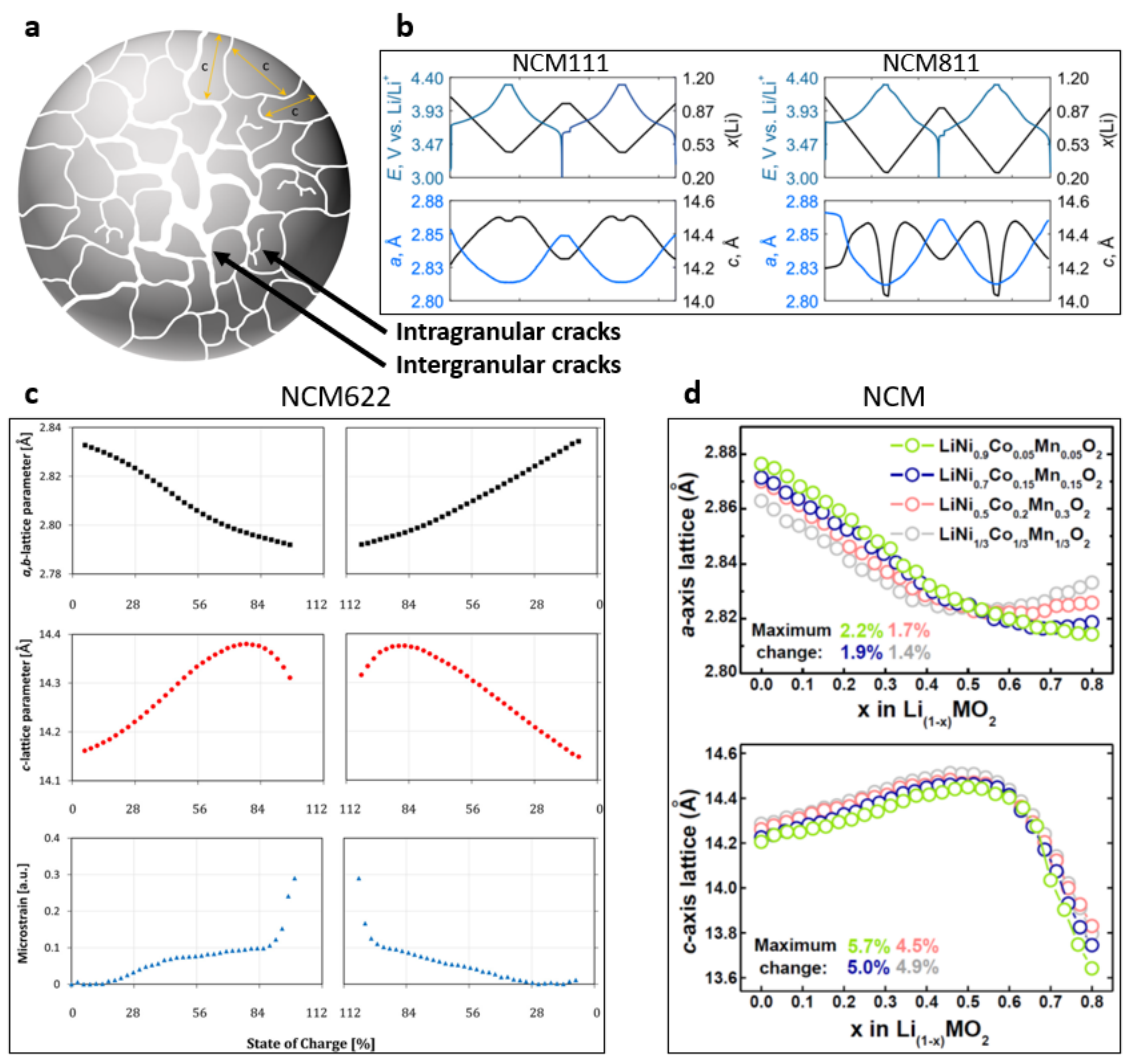

**Figure 3.** a) Illustration of intergranular and intragranular cracks and anisotropic lattice orientation in NCMs, b) changes in lattice parameters during charge-discharge in NCM111 and NCM811, reprinted with permissions from A.O.Kondrakov et al.[69], copyright 2017 American Chemical Society, c) changes in lattice parameters and microstrain during charge-discharge in NCM622, reprinted from ref.[73], licenced under CC BY-NC-ND 4.0, d) changes in lattice parameters during charge-discharge in NCMs with different stoichiometry, reprinted with permission from W. Li et al.[39], copyright 2019 American Chemical Society.

### Reordering of the layered structure – oxygen evolution and cation mixing

During battery operation, the stoichiometry of NCMs gets disrupted as lithium exits the material. This destabilizes the layered structure of NCMs, and at a certain degree of delithiation irreversible charge compensation reactions and structural changes start occurring. These irreversible processes are evidenced by the concurrent surface layer phase transition from layered to spinel to rock salt and oxygen evolution from the cathode material. Although oxygen redox in bulk NCM is reported mostly reversible[74], at high degrees of delithiation (over 80% SoC), considerable oxygen evolution from NCMs has been reported by using online electrochemical mass spectrometry (OEMS)[40,75]. Furthermore, it was shown by operando emission spectroscopy that oxygen evolves in the form of singlet oxygen $^1O_2$[76], which is very reactive and either undergoes quenching to form ground state triplet oxygen or attacks the electrolyte, evoking further gas release in the form of CO and $CO_2$[40,75,76]. In addition to causing chemical electrolyte oxidation, the evolved oxygen also leaves behind an oxygen-depleted host material which is destabilized and transforms from layered ($MO_2$) to lower-oxygen-content phases such as spinel ($M_3O_4$) and rock salt (MO). The presence of these oxygen-depleted phases in NCM after cycling has been detected by HR-TEM[77].

To investigate the simultaneous surface reconstruction and oxygen evolution in NCMs, it is important to understand the underlying charge compensation mechanism causing the evolution of $^1O_2$. Although current knowledge on NCM materials still includes certain ambiguity about the redox activity of the transition metals, it is generally agreed that Ni is the main redox active transition metal. It has also been widely reported that oxygen participates in charge compensation in NCMs by sharing electrons with TMs via hybridized TM-O orbitals[38,65,78–80]. A study from 2021 by Kleiner et al.[38] utilized near-edge X-ray absorption fine structure spectroscopy (NEXAFS) along with charge transfer multiplet (CTM) calculations to show that charge transfer in NCMs occurs mainly via $Ni^{2+}$ oxidation to $Ni^{3+}$ along with an increase in covalency of the Ni-O bond which enables indirect participation of oxygen in charge compensation. Only Ni and O were reported to be redox active in all NCMs. Furthermore, nickel oxidation to $Ni^{4+}$ was refuted in all NCMs due to inconsistencies between their obtained NEXAFS results and CTM calculations for $Ni^{4+}$. Moreover, little to no participation of Co in charge compensation was claimed.

Another study on the charge compensation mechanisms in NCM811 confirms no redox activity of Co and Mn up to 75% SoC; however, continuous Ni oxidation up to $Ni^{4+}$ was reportedly measured by hXAS[65]. Many computational and experimental XAS reports on NCM111 have stated otherwise, confirming Co participation in charge compensation above 66% SoC[81–83]. Whereas, yet another study states that Co redox is not part of the oxidation processes in NCM111 up to 5 V and, furthermore, Ni on the surface is oxidized to $Ni^{3+}$ and in bulk – to $Ni^{4+}$[78]. The different conclusions are likely due to challenges in interpreting the results of X-ray absorption spectroscopy, as the edge shifts may occur not only due to changes in the oxidation states of the probed ions but also due to structural and electronic changes in the ligand field[78]. Thus, further studies on the charge compensation mechanism, whether it involves Co and whether nickel oxidizes to $Ni^{4+}$ in NCMs, are needed to describe the mechanism of oxygen evolution exactly and to build a foundation for understanding other structural degradation mechanisms in NCMs. However, it is generally accepted that oxygen evolution originates from its redox activity and electron sharing with TMs via hybridized TM-O orbitals which allows the loss of electrons from oxygen. Hence, at high SoC (80%) an onset of oxygen evolution is observed[40,76].

In the already mentioned 2021 study by Kleiner et al.[38], it was reported that the Ni-O bond formed with the $Ni^{2+}$ species is highly ionic in character, whereas the Ni-O bond formed with $Ni^{3+}$ species is highly covalent, promoting electron sharing between Ni and O. Furthermore, Kleiner et al. found that all $Ni^{2+}$ is converted to $Ni^{3+}$ above 155 mAh/g capacity (56 % SoC) for NCM 111, 622 and 811 materials prepared in their study, after which irreversible degradation processes start occurring. After this point, a slight increase in the amount of $Ni^{2+}$ was reported. Thus, if indeed no $Ni^{4+}$ is formed and Co does not participate in charge compensation, it would mean that in the range up to 80% SoC, the electrons for charge compensation are ensured by redox reactions involving electron sharing between nickel and oxygen which do not oxidate the oxygen to ground state yet, as no major oxygen evolution is observed up to 80% SoC[40,76]. It would then follow that beyond 80% SoC some oxygen irreversibly oxidizes to a point where $^1O_2$ is formed ($O^{2-} \rightarrow \tfrac{1}{2} {}^1O_2 + 2e^-$) and evolves from the material, triggering the detrimental phase transition from layered to spinel and rock salt phases with decreasing amount of oxygen:

$MO_2$ (layered) $\rightarrow$ $M_3O_4$ (spinel) $\rightarrow$ MO (rock-salt)[76]

Oxygen redox occurs both in bulk and surface regions; however, due to kinetic reasons, higher SoC is being reached on the surface. Hence the involvement of oxygen in charge compensation on surfaces of cathode particles is much more pronounced and irreversible[84,85]. Consequently, more oxygen exits the active material, and more pronounced phase transformation is observed closer to the surface of the particle[40,86]. After the initial cycles, a growing oxygen-depleted surface layer is formed on top of the particle, lowering the amount of oxygen released in the following cycles, as the diffusion path of oxygen to the particle surface increases. This has been verified experimentally by R. Jung et al.[40] by cycling NCM111 and NCM622 up to 4.8V and NCM811 up to 4.4V for four cycles and measuring the gas evolution by OEMS. It was observed that the gas evolution is strongest in the first cycle and decreases in the subsequent cycles, verifying their hypothesis that oxygen diffusion from the bulk of the particles is limited given sufficient particle size and low to moderate temperatures. The hypothesis is further confirmed by numerous other reports[87–90] showing that the phase transition to spinel and rock salt phases in NCMs is limited to the surface regions. While the self-passivating behaviour is generally desirable, surface spinel and rock salt phases also hinder lithium diffusion increasing the cell impedance and leading to LAM.

Although no correlation has been found between Ni content in NCM and gas release from NCMs when charging up to 4.3 V, it was found by using OEMS that the release of $O_2$ and $CO_2$ gas was significant in the voltage range from 4.3 to 4.7 V vs. Li/Li$^+$, exhibiting a positive correlation with the Ni content[91].

In summary, delithiation to high SoC causes the oxygen to oxidize to its ground state and exit the material mainly from the surface regions along with surface reconstruction from layered to spinel and rock salt phase. The evolved reactive oxygen $^1O_2$ contributes to faster electrolyte decomposition, TM dissolution, and CEI growth[92]. In addition, the phase boundary between the surface rock salt or spinel region and bulk layered structure can accumulate stress during battery operation as the adjacent phases undergo varying degrees of changes in lattice parameters. This can result in intragranular cracks (loss of contact within primary particles), further impeding the rate capability and cyclability of the NCM cathode.

Cation mixing is considerably less pronounced in low-Ni NCMs (NCM111)[93,94] than in high-Ni NCMs (NCM811)[95,96]. The positive correlation between Ni content and cation mixing is due to proportionally less Co in the lattice of NCM811 which can mitigate the phenomenon[97], a more significant amount of $Ni^{2+}$ ions, which are prone to diffusing from the TM to Li layer due to similar $Ni^{2+}$ and $Li^+$ radii, and the lower voltage/faster rate at which critical SoC is achieved (80% SoC is reached at around 4.3 V for NCM811 and 4.6-4.7 V for NCM622 and NCM111)[37,40].

The I(003)/I(104) XRD peak intensity ratio can be used as an indicator of the degree of cation mixing[98], which increases, as the ratio decreases below a critical value of 1.2. Additionally, the (003) peak shift during prolonged cycling indicates structural instability[99]. However, precisely quantifying the degree of cation mixing, or the amount of Li in 3*b* and Ni in 3*a* sites, requires carrying out Rietveld refinement of XRD data[100]. Like with increasing voltage, it has been shown that cation mixing in NCMs increases with temperature[101]. This is explained by the increasing amount of oxygen vacancies upon heating, which facilitate transition metal (mostly Ni) migration from the TM layer to the Li layer.

It has been shown that cation mixing can be reduced by preparing NCM from Li-rich precursors creating a new class of Li-rich NCMs[93,102]. While this might be considered to be a separate class of materials, these can also be treated as Li-doped NCMs as some TM sites are occupied by Li instead of TMs. The properties of Li-rich NCMs notably differ from NCMs – it has been shown that oxygen participation in redox processes within the Li-rich NCM bulk does not lead to significant oxygen evolution. Rather, it is a reversible process throughout the material, greatly improving the specific capacity[103,104]. However, the high capacity comes at the cost of increased voltage hysteresis, sluggish kinetics, voltage fade, general stability, and cycle performance[34,105]. Furthermore, adding excess lithium during cathode material synthesis increases the undesirable RLC formation[106].

## TM dissolution

Any of the transition metals in the NCM cathode are prone to dissolution at high voltages and temperatures[107]. It has been reported that of the TMs present in NCM, the most detrimental effect on Li-ion battery operation is caused by Mn dissolution, as it can cause significant damage at the anode side[108,109]. $Mn^{2+}$ can form complexes with carboxylate groups from electrolyte solvent breakdown or react with HF from the reaction between the electrolyte and residual $H_2O$ and diffuse towards the anode to be deposited on its surface[53,110].

At voltages beyond 4.3 V, increased TM dissolution has been observed for NCM111[111] since NCM in a delithiated state is more receptive to corrosion by electrolytic degradation products[112]. It has been observed that all transition metal dissolution from NCM is also enhanced due to the presence of a larger amount of HF, with $Mn^{2+}$ being the largest amount of dissolved species within the electrolyte[108]. Additionally, the extent of TM dissolution is affected by the electrolyte salt used in a battery due to their varying resistance to ambient moisture and, thus, the amount of corrosive HF produced[108]. Reaction with HF creates $MnF_2$, which can easily migrate towards the anode or remain incorporated within the CEI. The metal concentrations within the aged electrolytes are most often determined by inductively coupled plasma optical emission spectroscopy (ICP-OES)[107,108,111]. TM dissolution from the cathode material can also be detected by probing the anode for trace amounts of TMs by X-ray photoelectron spectroscopy (XPS), X-ray absorption spectroscopy (XAS), and other techniques[6,113–116].

At low potentials (highly lithiated state), manganese dissolution follows the disproportionation mechanism[117]:

$2Mn^{3+} \rightarrow Mn^{4+} + Mn^{2+}$

In NCMs, Mn ions are predominantly in the 4+ oxidation state, which is electrochemically inert and stable[118]. However, some $Mn^{3+}$ may be present due to lattice defects and oxygen vacancies[119]. More oxygen vacancies are observed in NCM811 than in NCM111 and at lower potentials, suggesting that Mn dissolution could be more pronounced in Ni-rich materials.

When $Mn^{2+}$ ions migrate to the SEI on the anode side, they can be incorporated into the SEI, blocking lithiation pathways[120]. Furthermore, manganese on graphite anode gets electrochemically reduced, followed by oxidation via reducing the electrolyte solvent[113,121]. Such a detrimental redox cycle ensures constant loss of active lithium via reaction with $MnCO_3$ species present in the SEI[121]:

$2Li^+ + MnCO_3 + 2e^- \rightleftarrows Li_2CO_3 + Mn^0$

$Mn^0 + EC \rightleftarrows MnCO_3 + C_2H_4$

### Cathode-electrolyte interphase formation

Cathode-electrolyte interphase (CEI) is formed due to multiple RLC reactions (see reactions 2.-5. in chapter 2.2.) and the RLCs themselves[32,122]. Additionally, chemical and electrochemical electrolyte oxidation, along with $LiPF_6$ salt breakdown and its reactions with the active material and RLCs, contribute to CEI formation[123].

The main compounds found in CEI depicted in figure 4 are TM fluorides ($TMF_n$) formed by TM dissolution and HF reactions with the oxidized active material, RLCs ($Li_2CO_3$, $LiO_2$, LiOH) discussed above, and electrolytic decomposition products (organometallic compounds, polycarbonates, lithium hexafluorophosphate oxidation products, and other side products). As there are many sources of CEI formation, many dependent more on the electrolyte than the cathode, it is difficult to define a comprehensive mechanism for its formation. It depends on the kinetics of TM dissolution, electrolyte breakdown, the amount of RLCs and the nature of their reactions with the environment, and other side-processes which may not be directly linked to the growth of CEI (e.g., oxygen evolution).

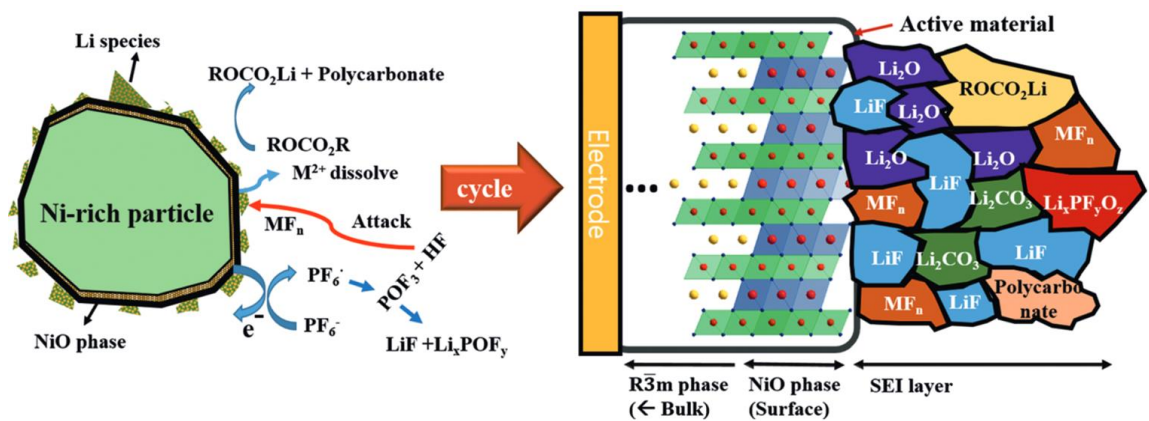

**Figure 4.** Evolution and composition of CEI, reprinted with permission from W. Liu et al.[32], copyright 2015 WILEY-VCH Verlag GmbH & Co. KGaA, Weinheim.

Since CEI is formed largely from lithium containing compounds, it contributes to LLI. The effect of CEI growth on the capacity of the battery is thus detrimental. Surely, transition metal dissolution (related to LAM ageing mechanism) also contributes to the growth of CEI, as TM compounds (oxides and fluorides), can be a part of the CEI[124]. While the growth of the CEI hinders lithium transport (and thus reduces the rate capability of the electrodes), it is self-limiting[125]. Since CEI grows in a uniform manner around the entire active material like a protective coating, like any surface coating it could be regarded as a passivating layer which can inhibit further reactions with the electrolyte and further LII and LAM. Thus, to improve the cycling stability, a stable CEI without compromising lithium inventory is an attractive solution. For this purpose, CEI-forming and CEI-stabilizing additives such as fluoroethylene carbonate or vinyl carbonate are often added to the electrolyte to control CEI formation and re-dissolution into the environment, greatly improving the capacity retention[126,127]. Note, however, that a formation of a stable CEI does not help to avoid structural changes and all of the issues associated with LAM.

### Degradation mechanism interplay

Determination of the most detrimental degradation phenomena in NCMs would require a thorough study in which each of the associated phenomena (cation mixing, oxygen evolution, lattice collapse and microcrack formation, TM dissolution, and CEI formation) could be isolated and their effect on capacity, rate capability, and cycling stability evaluated. However, such a study is practically impossible, as the degradation mechanisms are interconnected and the onset of one will trigger or accelerate another, which then provokes yet another. The main instabilities, and thus, onset of degradation, in the NCM cathode are caused by the disruption of the $LiNi_xMn_yCo_zO_2$ stoichiometry which occurs when lithium exits the cathode and migrates towards the anode upon charge. For example, the onset of oxygen evolution at 80% SoC is caused by excessive charge compensation by lattice oxygen due to lithium extraction from the cathode. The oxygen evolution then causes the lattice to become oxygen depleted and undergo phase transitions on the surface of the particles, which in turn intensifies mechanical stress within the primary particles, generating intragranular microcracks, impedes lithium diffusion, reducing capacity, rate capability, and cycling stability. In this chain of degradation effects, the trigger is lithium extraction which should not be mitigated if high capacity is desired. The immediate consequence of lithium extraction is charge compensation by Ni and oxygen. This effect can be altered by various doping strategies discussed in further chapters which increase the effective charge of oxygen and allow it to remain in the lattice. Thus, in this chain of degradation effects it is paramount to mitigate the oxygen evolution by retaining the oxygen within the lattice.

Another example of degradation interplay is the formation of CEI from RLCs, dissolving TMs and electrolyte decomposition products. In this case the mitigation is focused on removing the RLCs from the cathode particle surface after synthesis, preventing TM dissolution, and oxygen evolution and contact with the electrolyte. These objectives are often accomplished by washing the cathode after synthesis, doping, or coating the active material, as discussed in part 4.

Overall, a dominant degradation effect can either be experimentally deduced for each individual cathode composition based on in-situ and post-mortem material characterization and electrochemical measurements at certain cycling conditions, or identified based on how many other degradation mechanisms it could trigger (e.g. oxygen evolution).

## 2.4. Degradation Beyond the Cathode

Besides cathodic degradation mechanisms, other degradation processes occur within the battery cell, which are related to the electrolyte and anode and are worth at least a passing mention. For example, at high voltages, electrochemical electrolyte oxidation occurs. However, the typical operational voltage range of a battery cell is chosen so that electrolyte decomposition can be considered negligible compared to the chemical oxidation with the released lattice oxygen[40,128] (typically with an upper cut-off at 4.3 V). Release of the reactive lattice oxygen, which is the main contributor to electrolyte degradation at typical operational voltages, occurs concomitantly with surface phase transitions which are closely related to the amount of Ni in the NCM cathode. Thus, the chemical oxidation of the electrolyte is also closely associated with the amount of Ni in the NCM cathode.

Based on the results of J. Wandt et. al.[76], in addition to reactive oxygen evolution, most $CO_2$ evolution also comes from the chemical decomposition of the electrolyte, with $CO_2$ evolution due to electrochemical oxidation of $Li_2CO_3$ on the cathode surface being negligible. Furthermore, several types of protic species (HF, ROH, $H_2O$) may result from chemical and electrochemical electrolyte oxidation[40,129], which can react with the salt in the electrolyte, forming further HF by the previously stated reactions (6.-8.).

Cathodic degradation processes also cause damage to the anode, such as excessive SEI growth, TM dissolution and deposition on the anode, and Li dendrite growth which damages the stabilizing SEI on the anode and further traps active Li as the damaged SEI areas are reconstructed. The SEI layer is formed during the first couple charge-discharge cycles from lithium and electrolyte decomposition products and it is critical for stable battery operation, as it prevents continuous electrolyte decomposition at the anode. As such, a stable SEI is also imperative for cycling stability and capacity retention. An unstable and continuously degrading SEI would leech lithium from the battery for rebuilding and thus cause LLI and negatively impact the capacity retention.

While the growth of lithium dendrites is negligible when operating most battery cells at room temperature at a reasonable rate, Li plating and subsequent growth of dendrites can become the dominating mechanisms of battery cell ageing as a whole when operating cells below room temperature[130,131]. LAM is an issue that many newer generation silicon-containing anodes encounter[132]. However, as the ageing and degradation of the anode is beyond the scope of this NCM-focused review, for further information, the reader is redirected to more exhaustive review articles on this topic[133,134].

## 3. Material-specific degradation mechanisms

### 3.1. Dominant Degradation Mechanisms in NCM111

In NCM111 cathode, the transition metals are in atomic ratio 1:1:1. Thus, NCM111 contains a more significant amount of Co and Mn (1/3) than subsequent evolutions of NCMs (e.g., NCM622, NCM811, etc.), hinting at manganese dissolution as a possible issue. However, it has been reported that transition metal dissolution is a minuscule source of capacity fade in NCM111 cathodes[118]. It was calculated from dissolved TM mass that after 100 cycles up to 4.3 V, the capacity fade attributed to TM dissolution was only 0.1 mAh/g, which is ~0.06% of the practical specific capacity (160 mAh/g) of NCM111 cathode material. The most detrimental effect of the dissolved TM (especially manganese) is the degradation of anodic SEI[108,109,113,135].

Microcrack formation in NCM111 particles occurs to a considerably lower degree than in Ni-rich (x>0.8) cathode materials when charged to the same voltage (4.35 V) due to lower extent of delithiation and hence more minor changes of lattice parameters (see figure 3b for details)[136]. It was, for instance, observed that after 50 cycles, the intergranular cracks formed in NCM111 remained contained within the particle, suggesting that the lattice volume changes in NCM111 material are not severe enough to break down the secondary particle. However, severe intergranular and intragranular cracks appear, when NCM111 is charged beyond its typical operational potential to 4.7 V (80% SoC). Hence rather than being a function of operational voltage, cracking in NCM materials is perhaps best viewed as a function of lithium content or SoC[66].

RLCs left after synthesis, and structural and chemical changes during battery operation at sufficient voltages can bring about battery degradation[60], as also described in the previous section. However, NCM111 is among the materials most stable in air when compared to other NCM compositions and high-Ni NCMs in particular. NCM111 is also the most thermally and structurally stable of the reviewed NCM materials charged up to a certain voltage, displaying the best capacity retention[118]. It is unfortunately also the NCM with the lowest practical capacity as it gets delithiated to a lower degree than NCM622 or NCM811 when charged to the same voltage. This is strictly due to the higher redox potential of NCM111, as the achievable capacity when charged to the same SoC (80%) for low and high Ni NCMs is the same (220-222 mAh/g, calibrated to the formula weight of the specific NCMs)[39]. Capacity fade for NCM111 is accelerated by increasing cut-off voltage to 4.6 V (near 80% SoC) and the cycling temperature from 25°C to 60°C[7]. The extreme conditions evoke oxygen vacancies accompanied by reactive oxygen release. Under increased temperature, these vacancies can migrate within the bulk of NCMs, enabling oxygen release from deeper parts of the cathode material[137]. In low-Ni NCMs the temperature required for vacancy migration was shown to be higher than in high-Ni NCMs.

In NCM111, the ratio of I(003) and I(104) peaks is larger than 1.2, indicating a low degree of cation mixing[138,139]. However, increasing the temperature and voltage exacerbates this phenomenon, as discussed in section 2.3. At high SoC, the surface of NCM111 material gets highly delithiated, allowing for easier cation mixing and surface reconstruction. Suppose most capacity is accessed by charging the battery to high voltages (80% SoC reached only at 4.7 V). In that case, the surface of the material will transform to spinel and rock salt phase, increasing cell impedance. Thus, if NCM111 is charged to high voltages for increased energy density, surface reconstruction is highly pronounced, and the cycle life – rather poor.

Overall, as the microcrack formation in the cathode material when charged up to 4.3 V is not detrimental to the point of particle breakdown, it can be concluded that the main degradation in NCM111 occurs on the particle surface, including surface reconstruction from layered to spinel and rock salt phase, concurrent oxygen evolution, TM dissolution, electrolyte degradation, and CEI growth. However, if NCM111 is delithiated beyond its typical operational voltage at 60% SoC (4.3 V) to 80% SoC (4.7 V), abrupt volume change, substantial surface reconstruction, and microcracking take place.

### 3.2. Dominant Degradation Mechanisms in NCM622

Subsequent evolutions of NCM have seen increasing nickel content and improved gravimetric capacity that can be accessed reversibly. Thus, we view NCM622, where 60% of the TM content is Ni, as the next example in the overall trend. When capacity retention and thermal stability are considered, it shows one of the closest performances to preferred (or ideal) cathode properties (figure 1b). As such, NCM622 has, in recent years, been among the most popular choices for battery manufacturers focussing on the long-term cycling stability. NCM622 (polycrystalline) has been shown to be stable (80% of initial capacity) up to 1375 cycles when charged to 4.2 V[140]. NCM532, a close relative to NCM622 meanwhile, has been demonstrated to show 90 % of its initial capacity even after 4000 cycles in the voltage range 3.0 – 4.3 V[12]; however, it exhibits a lower initial capacity when compared to NCM622[49].

Compared to NCM111, NCM622 has a higher Ni content (60%) relative to the other TMs. As is commonly known – higher Ni content means lower thermal stability and capacity retention at lower voltages. Among the reasons for the worsened properties are larger amount of RLCs left after synthesis, increased degree of cation mixing, surface reconstruction, and oxygen evolution, larger volume change during operation, and exacerbated reactivity of delithiated particle surface causing electrolyte degradation and CEI formation.

Microcrack formation starts causing significant degradation in NCM with growing content of Ni (figure 5e-i) as both the lattice parameter and volume change during operation within the operating voltage range become more pronounced (figure 3b-c)[49,141]. Microcracking after 200 cycles in NCM622 has been observed by some[142], even when charging NCM622 up to only 4.2 V, which corresponds to approximately 65% SoC where lattice parameter change is not yet substantial (figure 3c). As shown in the study by Ryu et al.[141], no peak in the dQ/dV plot of NCM622 cells is observed (figure 5d), as opposed to NCM materials with higher Ni content (figure 5a-c). The peak in dQ/dV plots shown in figure 5a-d has often speculatively been linked with lattice parameter change and microcracking[49,68,141]. Additionally, Ryu et al. showed that the microcracks formed in NCM622 were mostly arrested before reaching the particle surface. Furthermore, it was revealed that most microcracking, if any, occurs within the first few cycles. As discussed in previous chapters (see chapter 2.3. for detail), microcracking is followed by surface reconstruction of the newly exposed areas and subsequent CEI growth. It can hence be inferred that surface and bulk degradation are contributing factors in the degradation of NCM622 when it is charged up to 4.3 V.

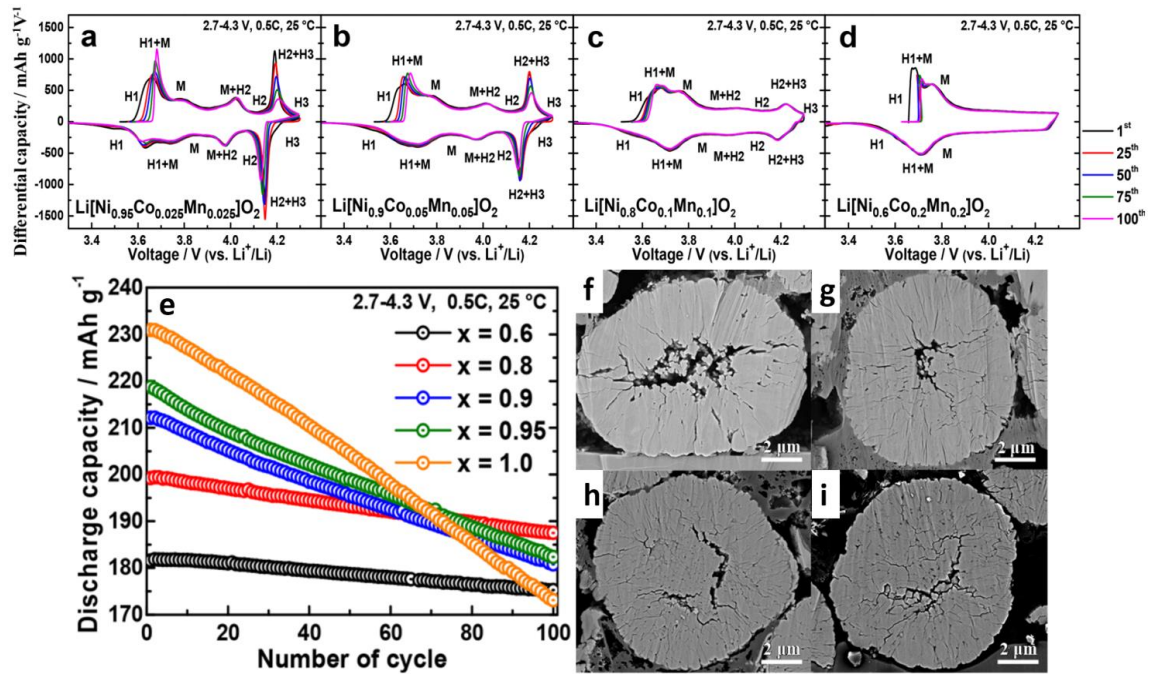

**Figure 5.** dQ/dV plots of a) NCM96, b) NCM90, c) NCM811, and d) NCM622; e) cycling curves of the respective composition cells; f-i) Cross-sectional SEM images of f)

NCM622, g) NCM811, h) NCM90, and i) NCM96 cathode materials in the fully charged state (at 4.3 V in the first charge), Reprinted with permission from H.-H. Ryu et al.[141], copyright 2018 American Chemical Society.

### 3.3. Dominant Degradation Mechanisms in NCM811

As the Ni content in NCM grows, so do the gravimetric and volumetric energy densities. Meanwhile, the reduced use of cobalt leads to the improved environmental sustainability of the material. Unfortunately, the stability suffers (figure 1b). As the landscape of the decomposition mechanisms becomes more complex at high SoC, more deliberate handling procedures and strategies for mitigating ageing are needed to protect the Ni-rich NCMs from premature decomposition.

In NCM811, the Ni composes 80% of TM content, suggesting that the properties of $LiNiO_2$-$LiMnO_2$-$LiCoO_2$ solid solution should be similar to the properties of the high-capacity $LiNiO_2$ component. It has been shown that $LiNiO_2$ undergoes an O3 → O1 phase transition (also denoted H2 → H3) at voltages around 4.2 V. This is represented by a peak in the dQ/dV plot, a plateau region on the charge-discharge curves, and has been proven by in-situ XRD showing clear (003) peak splitting at voltages around 4.2V[143,144]. Since NCM811 displays a similar anodic peak in the dQ/dV plot and a plateau in the charge-discharge curves starts appearing at 4.2V for x>0.8, a widely accepted misconception exists that a similar phase transition (H2 → H3) could be occurring for Ni-rich (x>0.8) NCMs[40,68,95].

Further exploring the phenomena, R. Jung et al.[40] found that the anodic peak in NCM811 at around 4.2V and the broad features around 4.6V for NCM622 and NCM111 are followed by a rise in oxygen evolution. Thus, it was believed that the H2 → H3 phase transition is accompanied by oxygen release. However, in studies from 2019 by K. Märker et al.[37] it was proven by using operando synchrotron XRD that no H2 → H3 phase transition occurs within the NCM811 cathode material. The anodic peak at 4.2 V is present and the lattice *c* parameter contracts, but in the XRD plots no splitting of the (003) reflection is detected. Furthermore, even when the splitting of (003) reflection is observed during the charging, the underlying reason is a non-homogenous distribution of lithium between the particles of the active material, as shown by W. Chueh's research group[145]. Thus, the widespread belief that the main degradation mechanism in NCM811 materials is the H2 → H3 phase transition is discredited, as no such phase transition occurs. Furthermore, it shows that estimating the presence of phase transitions based on the redox peaks in differential capacity plots, as shown in figure 6, is not an accurate approach for explicitly detecting phase transitions. Such contradicting claims call for additional evidence or, at the very least, caution when considering the existence of the H2 → H3 phase transition in NCM811.

If the phase transition is indeed present in LNO and absent in NCM811, then an in-depth in-situ XRD study should be conducted on Ni-rich NCMs with increasing Ni content to find at which composition the (003) peak splitting returns in equilibrium conditions, indicating an onset of the H2 → H3 phase transition. This would aid in determining the maximum possible amount of Ni content in NCM cathode materials to increase the capacity, simultaneously avoiding the onset of the highly detrimental H2 → H3 phase transition present in $LiNiO_2$. Additionally, the origin of the detected oxygen evolution increase, which follows the dQ/dV peaks for NCM811, NCM622, and NCM111 materials at respective voltages, should be fundamentally studied to understand the source of the redox peak and implement the most effective mitigation strategies.

The peak in the dQ/dV plot at 4.2-4.3 V (figure 5) is immediately succeeded by oxygen evolution[40]. As the peak does not originate from the widely believed H2 → H3 phase transition, which apparently does not occur in NCM811, it is most likely an indicator of one or more of the other degradation phenomena co-occurring with the peak. As surface reconstruction and related oxygen evolution is an effect that diminishes after the first few cycles, the connection of these degradation mechanisms with the peak in the dQ/dV plot is substantiated by the decreasing intensity of the dQ/dV peak in subsequent cycles. The connection between the dQ/dV peak and surface reconstruction could be tested by purposely introducing cracks in the cathode particles and investigating whether the reconstruction of the newly exposed surfaces provokes an increase in the dQ/dV peak intensity.

The collapse of the *c* lattice parameter initiates at around 4.04 V (50% SoC) and reaches values below those in the delithiated material at around 4.35 V (80% SoC)[69]. The significant volume change accompanying the lattice parameter contraction causes strain and microcracking within secondary particles. Microcracking is especially pronounced at high voltages or SoC and large current densities[99] and is typically accompanied by oxygen release[40]. Oxygen can either get trapped within the cathode particle, increasing the porosity, or escape to the electrolyte, causing cell bloating and accelerated electrolyte decomposition.

The causes for microcracking, however, are manifold. While surface reconstruction and subsequent oxygen evolution are likely a part of the problem (intragranular cracks), so is the expansion and contraction of the crystal lattice (intergranular cracks). During charging, the *a* and *c* parameters of NCM811 expand and contract unevenly (figure 3b)[65]. The *a* parameter undergoes constant shrinkage during delithiation, however, the *c* parameter (perpendicular to the layers) initially expands and then contracts sharply as the delithiation continues. While the trend is similar to NCM111 and NCM622, the magnitude of the parameter change is considerably larger for NCM811 at lower voltages. It can thus lead to significantly more changes in the microstructure of the particles. The *c* parameter is a combination of Li layer thickness and TM layer thickness. Since the radius of Ni decreases upon delithiation ($Ni^{2+}$ gets oxidized to $Ni^{3+}$, higher valence cations are smaller) and the covalency between Ni and O increases, the TM layer continuously shrinks. The initial expansion of the *c* parameter occurs due to an increase in the repulsion of the oxygen layers as Li exits the material and no longer shields the negative anions from each other, thus, the Li layer thickness increases. The collapse of the *c* parameter observed as the NCM811 is charged further, initially occurs due to continued decrease of TM layer. Starting at around SOC ≥ 75%, the shrinking is thought to be due to notable oxygen participation in charge compensation, which lowers the negative charge of the oxygen layers, decreasing their electrostatic repulsion and decreasing the Li layer thickness considerably[37]. Moreover, extracting the capacity beyond 4.3 V (80% SoC) is difficult. It has been shown that along with the collapse of the *c* parameter, Li ion mobility also decreases, making it more difficult and perhaps even futile to further delithiate NCM811[37].

The effect of increasing Ni content on the generation of microcracks in NCM is clearly visible in figure 6. The cracks formed after 50 cycles up to 4.35 V are notably more pronounced in NCM811 (figure 6g-h) than in NCM111 (figure 6c-d) due to more extensive lattice parameter change occurring for NCM811 than for NCM111 at this voltage range. Furthermore, whatever little cracks have formed in NCM111 are mostly contained within the particle bulk, whereas the cracks formed in NCM811 propagate to the surface of the particle, exposing the bulk to electrolyte and secondary degradation mechanisms such as electrolytic degradation, CEI growth, TM dissolution from the bulk, and reconstruction of the newly formed surface. The same trend can be seen from a study on a related material – NCA[146], where, in addition to microcrack formation, the extent of capacity fade with increasing Ni content is shown. The material with the least amount of Ni (NCA80) shows slight microcracking and a capacity retention of 80.6% after 1000 cycles to 4.2V. Whereas the material with the highest amount of Ni (NCA95) shows complete particle degradation after 1000 cycles and capacity retention of 17.5%.

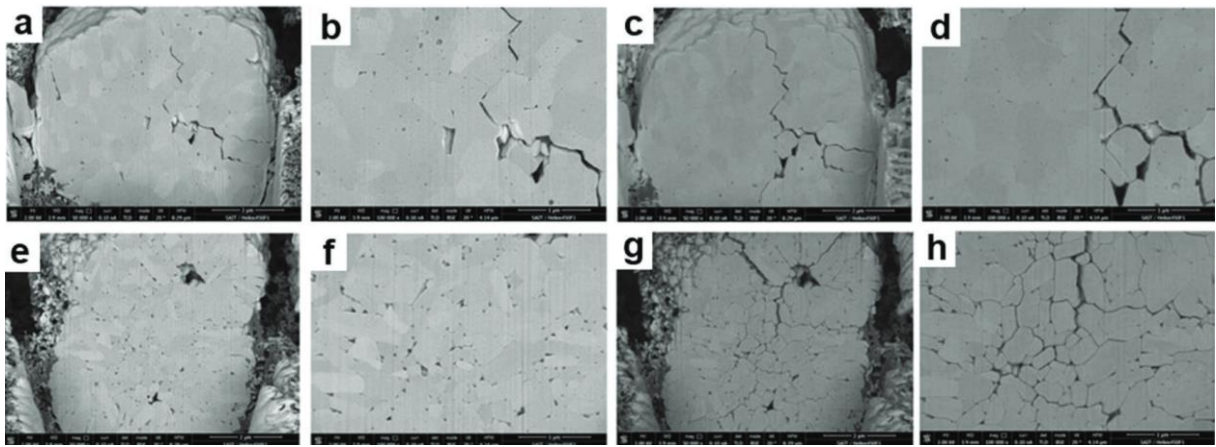

**Figure 6.** Microcrack distribution in NCM111 (a-d) and NCM811 (e-h) before (a-b, e-f) and after (c-d, g-h) being charged to 4.35 V for 50 cycles, reprinted from ref.[136], licenced under CC BY 4.0.

Lastly, NCM811 exhibits the largest amount of RLCs left on the cathode surface after synthesis. Thus, reactions involving the cell environment and CEI growth due to RLCs are also more pronounced than in lower Ni cathodes, degrading the cathode surface.

In summary, the surface degradation described for NCM111 and NCM622 is even more aggravated in NCM811. Most importantly, however, a highly detrimental lattice collapse comes into play as SoC exceeds 80% even at voltages as low as 4.3 V. While this allows obtaining higher practical capacity, it also causes particle disintegration and exposes a larger surface area for detrimental surface reactions. Thus, besides preventing surface degradation, the focus should also be on limiting microcrack formation by stabilizing the lattice and arresting the emerging microcracks before they reach the surface and cause particle disintegration.

### 3.4. Summary of degradation in NCMs based on Ni content

In NCMs with varying Ni content the dominant degradation mechanism is largely dependent upon how far the structure is delithiated. The critical 80% SoC is achieved at lower voltages for NCMs with increasing Ni content. Degradation in NCMs can be evaluated considering the most commonly applied upper voltage threshold of 4.3 V vs. Li/Li$^+$ for NCM batteries, which is chosen keeping in mind the electrochemical stability window of the electrolyte and cathode materials. When charging the NCM battery up to 4.3 V, the dominant degradation mechanisms gradually shift from mostly surface based degradation in NCM111 to bulk degradation overshadowing surface deterioration in NCM811 as illustrated in figure 7. When comparing the cycling stability of NCM111 and NCM811, the overall degradation of NCM811 is much quicker than that of NCM111, however, a higher capacity can be achieved with growing Ni content, thus justifying the increasingly complex efforts devoted to improving the stability of high Ni cathodes.

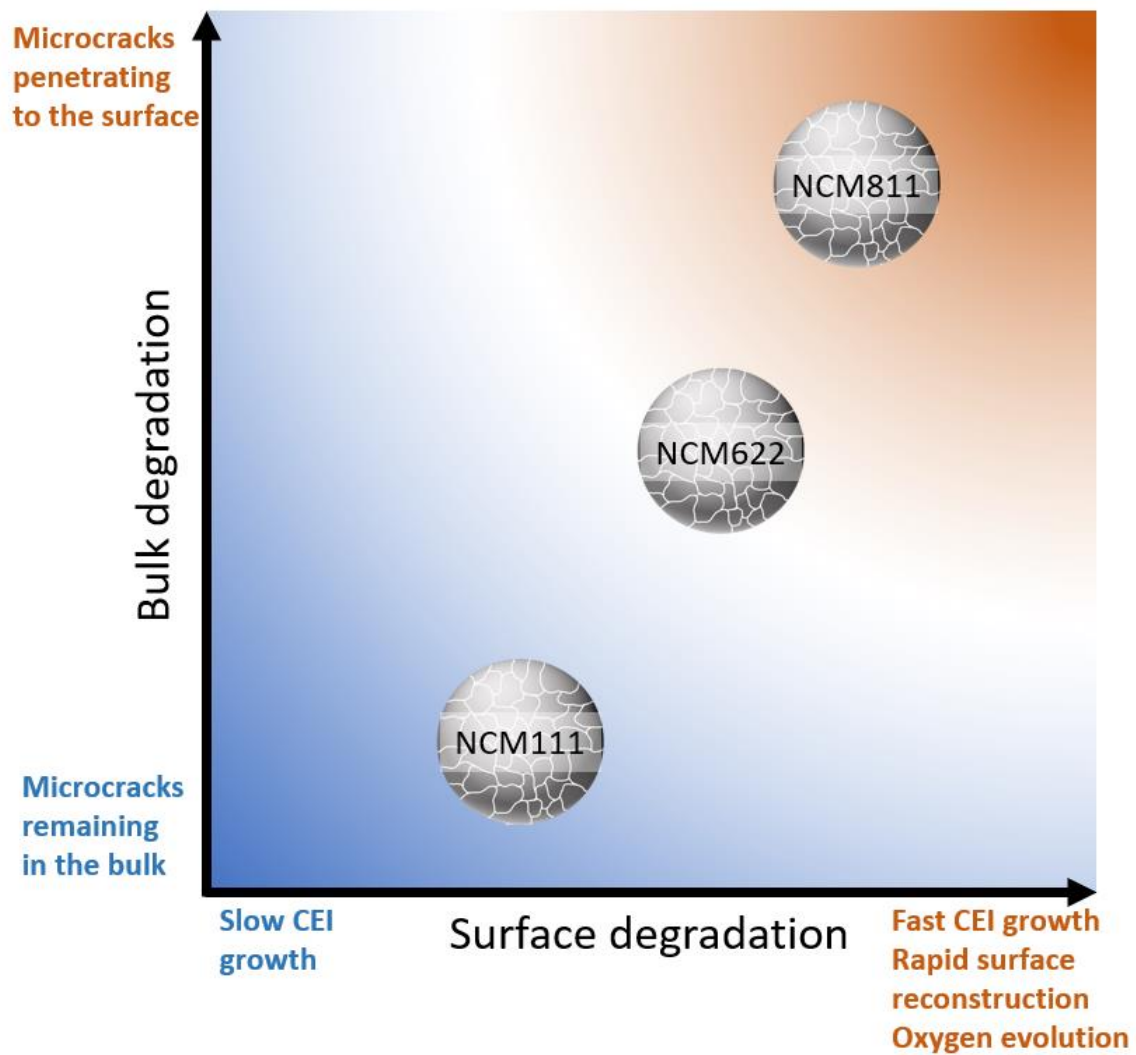

**Figure 7.** Illustration of the main type of degradation occurring in the NCM111, NCM622, and NCM811 cathodes during battery operation.

## 4. MITIGATION STRATEGIES

Several mitigation strategies have been proposed to counter the different degradation mechanisms occurring within NCM cathodes. The most prominent ones are cathode particle surface coating, doping, and the control of the particle microstructure during the synthesis of cathode particles (figure 8). In addition to improving the cycle life of cathode materials, the extent of improvement by different mitigation strategies may also give an insight into the dominant degradation mechanisms. If the effect of a certain modification strategy is known (e.g., doping improves structural stability and mitigates phase transitions, coating prevents surface degradation and hinders reactions with the electrolyte), then the extent to which it enhances battery stability can indicate the extent of detrimental effects from respective degradation processes.

In essence, the goal of cathode material modification is to improve the stability during electrode processing and cycling of the battery cell so that materials with high specific energy (NCM811, NCM83, Li-rich NCM) could also display good cycle and service life. This is especially important because the cyclability of many Ni-rich NCM materials is usually inherently poor (under 1000 cycles until 80% of the initial capacity is reached), which is detrimental to their entry into the market.

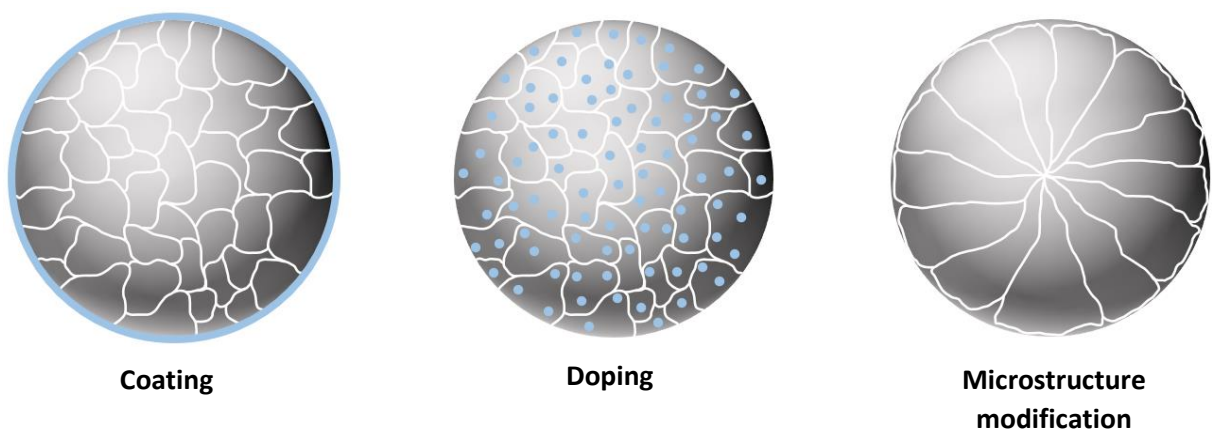

**Figure 8.** Illustration of the main cathode degradation mitigation strategies – coating, doping, and microstructure modification.

The different approaches to enhancing cathode cyclability (figure 8) all target different degradation aspects of the cathode particle, so that they can be matched to the observed degradation mechanisms for maximum improvement in cycle life. Here, we focus on materials development-related improvements in external cycling conditions or overhauling of the underlying chemistry of the active material, and hence we identify surface coating, doping, and structural control during synthesis as main pathways towards more stable NCM materials. Beyond the scope of this paper remain cell-level improvements that could benefit the improvement of cycle life, such as improving electrolyte, tailoring of electrode architecture, optimizing operational temperature, adjustment of negative to positive electrode ratio (N/P ratio) of the cell, limiting the current and voltage range, etc.

### 4.1. General mitigation strategies

#### Surface coating

Surface modification of cathode particles can be done by coating the particles or whole electrodes with a protective layer or doping the surface region with foreign ions. Several methods have successfully been implemented for coating NCM particles – atomic layer deposition (ALD)[147–149], chemical vapor deposition (CVD)[150,151], magnetron sputtering[152], and wet-chemical synthesis[153–156] are amongst the most popular. Both electrodes and active material particles can be coated, each having its own advantages (more uniform coating on all particles vs. improved electrode kinetics). While ALD, sputtering, and CVD techniques have, to some extent, all been upscaled successfully, the chemical approach is by far the simplest path that can be up-scaled more cost-effectively for commercial applications. Nevertheless, the uniformity of the coatings is largely better when deposited by ALD or CVD, with the magnetron-sputtered and wet-chemically-coated materials often displaying non-uniform coating.

The aim of coating the surface of a cathode particle or electrode is to create a thermodynamically and chemically inert layer on the particle without increasing the internal resistance of the battery too much. This way, the cathode can be protected from undesired side reactions with the environment, inhibiting CEI formation. Surface coating can also prevent the loss of active material by hindering oxygen evolution, TM dissolution, and phase transitions. Some of the most researched cathode coatings are electronically conductive carbon coatings[157], metal oxide ($Al_2O_3$[147,156,158,159], ZnO[160,161], $ZrO_2$[156,159], $TiO_2$[159,162], MgO[149,163] or rare earth oxides[164,165]) and Li-containing ionically conductive coatings such as $Li_3PO_4$[153], LiF[166], $Li_3VO_4$[167], glasses ($Li_2O$-$2B_2O_3$[168], glassy phosphates[169], etc.) or organic polymer coatings[170,171]. Many inorganic coatings can be implemented in amorphous and crystalline forms. For an extensive review of the coating materials, the reader can be referred to one of the most recent coating-focused reviews[19,172–174].

One way to classify coatings is based on their material properties (figure 9). Among the most relevant properties for proper functioning of the cathode are electronic (figure 9a) and ionic conductivity (figure 9c). Thus, coating materials are often selected on these bases, however, frequently materials with low electronic and ionic conductivity (figure 9b) are also selected if they present other performance-enhancing attributes, such as good mechanical properties (hardness, elasticity), HF-scavenging ability, or chemical inertness.

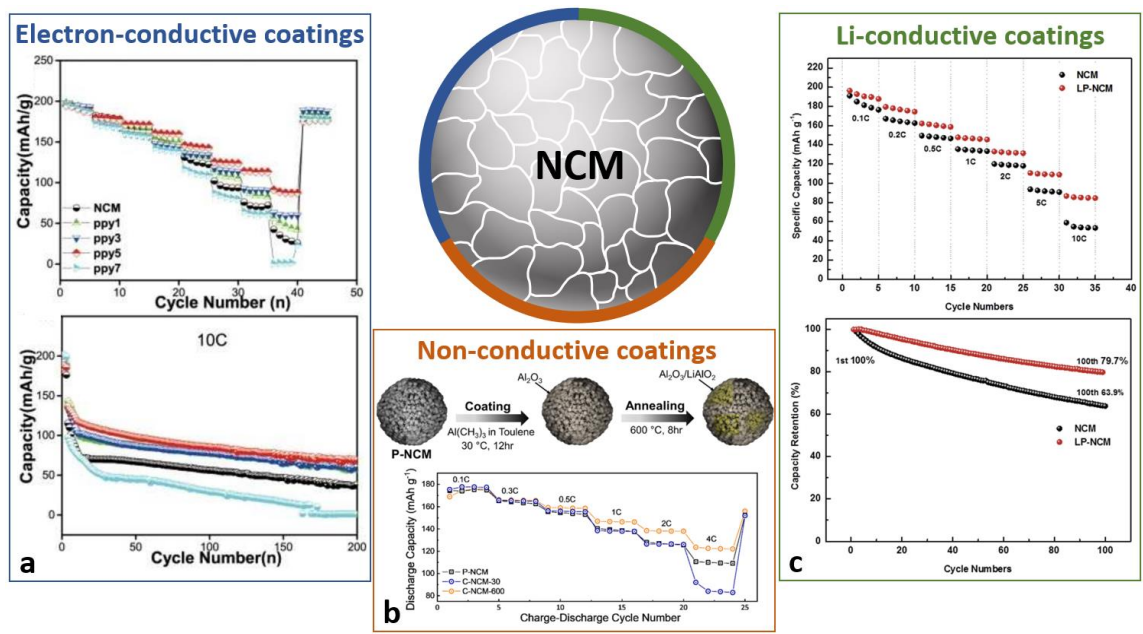

**Figure 9.** Types of coatings on NCM and their effect on the electrochemical properties. a) Effect of polypyrrole coating on the rate capability and capacity retention of NCM811, reprinted with permission from B. Li et al.[171], copyright 2019 Wiley-VCH Verlag GmbH & Co. KGaA, Weinheim, b) effect of $Al_2O_3$ and $Al_2O_3/LiAlO_2$ coating on the rate capability of NCM701515, reprinted with permission from R. S. Negi et al.[175], copyright 2021 American Chemical Society, c) effect of a $Li_3PO_4$ coating on the rate capability and capacity retention of NCM622, reprinted with permission from S.-W. Lee et al.[176], copyright 2017 Elsevier.

*Electronically conductive coatings*

Carbon-based coatings, including amorphous and graphitic carbon formed from various carbon-based precursors[157] as well as more complex compounds, such as graphene oxide or reduced graphene oxide[177], have traditionally been used to improve the conductivity as well as prevent surface-based reactions (Mn dissolution, HF etching, etc.). Another class of materials studied for this purpose has been conductive polymers with the advantage of forming a continuous, ultrathin layer of conductive material over the cathode particles with the ability to suppress unwanted reactions with the electrolyte[171,178–180]. Indeed, Xiong et al.[178] reported greatly improved rate capability and cycling stability, both at room temperature and at 60°C, for NCM coated with conducting polypyrrole. Similar findings have been reported by Li et al.[171] (figure 9a) and elsewhere[170,180–182]. In addition to protecting the cathode surface against unwanted side reactions, a polymer coating often displays great flexibility, allowing the contractions and expansions of the NCM lattice without tearing or cracking[180,181].

*Ionically conductive coatings*

In Li-ion batteries Li$^+$ ion conductivity is one of the key parameters ensuring good rate capability and capacity. Although most Li-containing coatings ensure decent Li$^+$ diffusion, their poor electrical conductivities impose thickness limitations, and hence most coatings have a thickness of only a few nanometers. As demonstrated by Ni et al.[183], the cycling stability of NCM was improved by a 1% (mass ratio to NCM) Li$_2$ZrO$_3$ coating. However, increasing the mass ratio of Li$_2$ZrO$_3$ beyond 1% resulted in a decrease in the capacity. The improvement to cycling stability was due to the chemical inertness of the coating, which protected the surface of the cathode particles from reactions with the electrolyte. The decrease in capacity when the mass ratio of Li$_2$ZrO$_3$ increased was possibly due to the thickness of the coating inhibiting electron transport.

Li$_3$PO$_4$ coating synthesis has been widely investigated due to the simple and beneficial fabrication route[153,174,184,185]. The most popular method is synthesizing the coating via a wet-chemical procedure by adding phosphoric acid. This makes use of the residual lithium compounds (LiOH and $Li_2CO_3$) to form the $Li_3PO_4$ coating. Not only does it mitigate the negative effects of RLCs discussed previously (chapter 2.2), but it also improves $Li^+$ diffusion due to being a good Li ion conductor and protects the surface from detrimental side-reactions. Jo et al.[184] and Lee et al.[176] (figure 9c) have reported a $Li_3PO_4$ coated NCM622 with notably improved rate capability and cycling stability, confirming these improvements.

*Non-conductive coatings*

Non-conducting coatings are applied to improve the mechanical and chemical stability of cathode materials, primarily to mitigate degradation. A very important aspect of consideration when synthesizing non-conducting coatings is the thickness. If the coatings are thick, they will protect the cathode material from degradation very well, however, the electronic and ionic conductivity will suffer considerably, as well as the gravimetric capacity. Thus, a compromise between degradation mitigation and conductivity inhibition should be achieved by synthesizing optimal thickness coatings.

As was suggested by Jung et al.[186], the discharge reaction of NCM with an $Al_2O_3$ coating firstly proceeds through lithiation of the coating to form $Li_xAl_2O_3$ and only the remaining lithium enters the NCM layers. Thus, a reduction in capacity is expected. Indeed, increase in $Al_2O_3$ layer thickness has been shown to reduce the initial capacity of NCM[187,188]. However, the use of alumina coating was outweighed by the improvements in cycling stability.

A method to improve the conductivity of alumina coating was reported by Negi et al.[175] (figure 9b). Post-annealing of the coated cathode material was carried out at 600°C for 8 h. Such heat treatment triggered lithium diffusion into the $Al_2O_3$ layer to form an $Al_2O_3/LiAlO_2$ coating with considerably improved lithium ion conductivity.

Similar to $Al_2O_3$, $TiO_2$ coating shields NCM from side reactions with the electrolyte and improves cycling stability. Fan et al.[189] reported reduced initial capacity of $TiO_2$ coated NCM, however the overall cycling stability was greatly improved.

If the coating has a higher mechanical strength than the cathode particle (e.g., amorphous $Al_2O_3$ hardness up to 13 GPa[190], NCM111 hardness 11.2 GPa[191]), the mechanical properties are also improved by coating – the coating can hinder extensive particle volume change and delay microcrack formation. Simultaneously, the coating must allow good Li ion and electron diffusion not to increase the impedance of the battery and ideally be made from abundant materials. Thus, the thickness of non-conductive coatings should be strictly controlled.

*Other coating effects*

Two other less-discussed benefits of surface coatings are the possible extension of the electrochemical window of the electrolyte and the improvement of Li-ion transport due to interfacial phenomena. Firstly, CEI, artificial (as in the case of coatings) or "natural" (formed during the operation, as in the case of CEI growth in the cell during cycling), can mitigate the Li chemical potential discrepancy between the electrolyte and electrode at the interfaces[192]. Thus, the electrochemical window can be extended due to the additional drop in chemical potential that the thin CEI layers provide. Secondly, a space-charge layer is typically formed between two materials with different chemical potentials in contact with each other where charge neutrality cannot be established by electron or atom migration. It has been shown that this affects Li-ion diffusion[193] and the transport of ionic species in general[194]. The space-charge effect can increase ionic diffusion in solid-solid dispersions[195] or even ensure additional mass storage[196], but it can also have a detrimental effect on the conductivity of the cell[197]. While the effects of the space charge layer can range from negligible[198] to considerable[199], it is one of the phenomena that needs to be considered in the broader discussion of protective coatings.

Chemical coating procedures include a final calcination step to create the desired layer on top of the cathode particle. As has been shown for $ZrO_2$ coating on NCM, annealing at temperatures >700°C causes $Zr^{4+}$ migration into the bulk, effectively doping the outer layer of the cathode particle[200]. Conversely, for $Al_2O_3$ coatings, no $Al^{3+}$ diffusion into the bulk of NCM is observed even at annealing temperatures as high as 800°C[187,201], explained by magnetic coupling of TMs which inhibits aluminium diffusion into the particle. Thus, a thorough analysis of the distribution of coating elements within the cathode particles should be conducted to differentiate between the effects of just coating and coating accompanied by doping. Unfortunately, most coating studies lack a comprehensive approach, rendering their results ambiguous regarding the reason (coating vs. doping) for the improved electrochemical performance. Thus, it can be difficult to separate the effects of coating vs. doping to ascertain which approach brings the most benefits for cycle life improvement. However, the general effects of coating (surface protection) and doping (bulk lattice stabilization) have been widely agreed upon. To achieve both surface and bulk protection simultaneously, several studies also report coating and doping NCM material[185,202,203]. However, often these studies include separate steps for coating and doping or require separate source materials for doping and coating. If both modifications can be achieved via a single modification route using the same source material, it becomes more time- and cost-efficient, while achieving the same goal of structural stabilization and surface protection.

## Doping

Doping is an often-used method in materials development where one or several elements are introduced in the target structure, thus altering its properties. Doping refers to an impurity effect, which influences the defect structure but not the ground structure and is hence usually limited to low concentrations[204]. The phenomenon has been reviewed in great detail in the recent scientific literature[173,204–206]. While a lot of progress has initially been achieved with doping in the semiconductor industry, it has become a widely used technique in materials development in general. Thus, it is a relatively typical practice to consider doping for cathodes of Li-ion batteries, where typically part of the transition metal atoms is substituted by another element to change the electronic structure and crystallographic parameters, usually to alter specific aspects of the material performance. Many empirical experimental and theoretical works have been focusing on studying doping for NCM, NCA, LFP, HV-spinel, and other electrodes for Li-ion batteries[173,207]. They are discussed further on in the text.

There are two possible pathways for improvements in electrochemical performance and cycle life resulting from doping – improving conductivity of the material (electronic and/or ionic) and mechanical stability. The changed conductivity results from the introduction of additional electrons or electron vacancies, thus altering the charge carrier concentration, band structure, and overall conductivity of the material. The other effect of doping is mitigating the lattice changes during (de)lithiation. The valency and size of a transition metal ion influence its bond length and ideally can mitigate the volume change that results from Li insertion/extraction in the host structure. While in the case discussed later for NCM materials doping generally seems to be beneficial and mostly mitigates the volume changes, the case of another cathode LiFePO$_4$ is somewhat cautionary[208] – due to the antagonistic role of ionic and electronic carriers and the inherently high stability and low volume changes of LiFePO$_4$, improvements in conductivity are limited or counterproductive[209–211]. Hence particle size reduction might be a more suitable strategy for improving the electrochemical performance of LiFePO$_4$[205].

The case of NCM, with 3 types of transition metals and a multitude of ordered and disordered phases[5], however, is arguably more complex and less understood. Moreover, the conductivity, in general, is higher for NCM materials[212] than for LiFePO$_4$[213]. Thus, the main improvements from doping of NCM stem from preventing the transition metal atoms from forming anti-site defects and mitigating the lattice contraction/expansion. A statistical literature survey presented in the subsequent chapters indicates good promise for the doping methods, especially in Ni-rich NCM.

In NCM cathodes, three general sites can be substituted by a dopant atom – lithium (3a), oxygen (6c), and TM (3b) site (figure 10). Some of the studied dopants include Al[214–220], Ti[219,221–224], Mg[215,219,225–227], Ta[216,228], F[229–231], polyanions[232–234], and others[219]. Based on the valency and ionic size of the dopant and the method of preparation it will inhabit one of these three sites. Doping the TM or Li sites with atoms that have stronger oxygen bonds or doping oxygen sites with F to create stronger TM-F bonds increases the structural rigidity[179,229,235]. It can prevent extensive lattice parameter change during battery operation, mitigate oxygen evolution and cation migration, thus slowing down surface reconstruction and rapid electrolyte deterioration. Additionally, simultaneous Li and O site doping has been shown to enlarge the interlayer spacing, thus improving Li diffusion and rate capability[235,236].

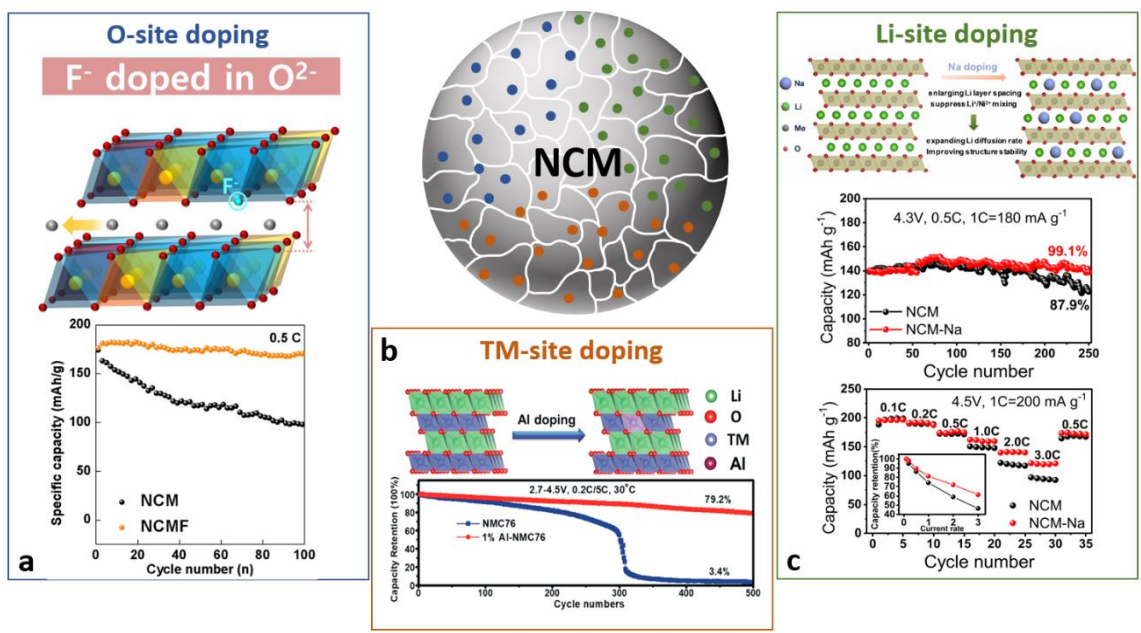

**Figure 10.** Doping sites in NCM and their effect on the electrochemical properties. a) Effect of F doping in O sites on the capacity retention of NCM811, reprinted from ref.[229], licenced under CC BY, b) effect of Al doping in TM sites on the capacity retention of NCM761014, reprinted with permission from W. Zhao et al.[237], copyright 2020 American Chemical Society, c) effect of Na doping in Li sites on the capacity retention and rate capability of NCM600535, reprinted with permission from Y. Shen et al.[238], copyright 2021 Elsevier.

*TM site doping*

TM site doping (figure 10b) is one of the most widely studied doping strategies for improving NCM performance[179,219]. TMs are usually substituted with other TMs which might stabilize the lattice and help reduce oxygen evolution. Doping the TM sites with atoms that have stronger oxygen bonds increases the structural rigidity preventing extensive lattice parameter change during battery operation and thus improving the cycling stability[219]. Introducing ions whose orbitals overlap to a higher extent with oxygen and which have higher charge-transfer capabilities can also hinder oxygen from exiting the lattice[219,239]. The charge transfer to oxygen ensures higher effective oxygen charge and thus, a higher binding energy, keeping the oxygen tightly bound to the transition metal. Hence, since oxygen evolution and phase transitions in NCM are concurrent, it follows that phase changes are mitigated by dopants increasing the oxygen bond strength as well. It has been reported that doping NCM90 (Li[Ni$_{0.90}$Co$_{0.05}$Mn$_{0.05}$]O$_2$) material with 1.0 mol% boron fully mitigates microcracking after 100 cycles at 55°C[240]. Boron doping altered the primary particle distribution forming radially organized secondary particles, which allowed the volume changes associated with (de)lithiation to be coordinated in a way to minimize intergranular stress. However, it was reported that the rate capability of B-doped material was slightly reduced compared to the pristine material.

To mention one of the more recent examples, Zhang et al.[72] recently reported a layered transition metal oxide with the formula LiNi$_{0.8}$Mn$_{0.13}$Ti$_{0.02}$Mg$_{0.02}$Nb$_{0.01}$Mo$_{0.02}$O$_2$ that shows almost no volume change upon delithiation and an excellent cyclability (85 % capacity retained after 1000 cycles at C/3, Li counter electrode exchanged in 500$^{th}$ cycle). In this work, the careful selection of transition metals helped avoid the change of lattice volume changes upon delithiation.

Another dopant which is widely studied and inhabits the TM sites (preferentially Ni site[239]) is Al. Dixit et al.[239] showed that strong Al-O iono-covalent bonding could stabilize the cathode material structure and improve charge transfer to oxygen, although slightly inhibiting lithium diffusion. Similar structural stability improvement by doping Al in TM sites has been reported by Li et al.[220] In the same study the case of doping Al in Li sites, however, brings about different effects to NCM performance.

*Li site doping*

Some dopants which inhabit the Li sites (figure 10c) in NCMs are K[223,241], Na[238], Mg[242], and other cations[228,243], based on the preparation method, such as the aforementioned Al$^{3+}$ [220]. The most common aim of doping Li sites is to improve the lithium diffusion kinetics and prevent the lithium layer collapse during charge-discharge cycling. The dopant ions are introduced as anchor points or pillars which are rigid and hold the lithium diffusion layer at a constant width[243]. In the study by Li et al.[220] aluminium doping in Li sites induced $LiAlO_2$ phase formation, which, in minuscule amounts, improved lithium diffusion kinetics. Larger amount of Al substitution in Li sites, however, had a detrimental effect on the cycling performance due to incompatibility between $LiAlO_2$ and NCM phases.

Doping Li sites of NCMs with alkali metals such as K or Na has been reported to achieve the aim of increasing the width of Li layer and improving Li diffusion, while simultaneously decreasing cation mixing due to increased energy barrier for $Ni^{2+}$ migration[238]. Mg and Ta doping in lithium sites of Ni-rich layered cathodes, on the other hand, proved to hinder lithium diffusion and lower the capacity[228,242]. The stabilizing effect of Mg or Ta pillars in the Li layer, however, did have a positive effect on the cycling performance.

*O site doping*

O sites (figure 10a) are doped by other negatively charged ions, such as F[244,245], Cl[246], S[245], or polyanions ($(PO_4)^{3-}$, $(SO_4)^{2-}$)[232,247]. While the aim of doping oxygen sites is to improve the stability and the overall performance of layered Ni-rich cathodes, in some cases the initial capacity is negatively impacted. For example, in many cases F doping has indeed been shown to improve the capacity retention, rate performance and thermal stability of NCM, however, the initial capacity is lower than for the pristine material[244,248–250]. Improvements to the capacity retention are commonly ascribed to fluorine protecting the NCM from HF attack. In the case of another halogen, chlorine, the initial capacity of NCMs is not diminished upon doping, and the capacity retention is improved as well[246]. Similarly to fluorine, the larger electronegativity of chlorine, compared to oxygen, ensures stronger bonds with the TMs, improving the stability of NCM.

Polyanion doping is not as popular as mono-ion doping. Residual sulphates, for example, are often deemed harmful to the performance of NCMs, however, improvements to both rate and cycling performances have been reported when purposely adding sulphate to the NCM precursor[247]. Sulphate doping was shown to reduce Li$^+$ diffusion barrier, simultaneously hampering the lattice contractions of NCM. Phosphate doping was shown by Cong et al.[232] to mitigate cation mixing and concurrent oxygen evolution in addition to increasing the c lattice parameter, which resulted in a higher cycling stability and rate capability, respectively.

*Multiple site co-doping*

In addition to doping one site and dopant at a time, multiple studies have shown that co-doping NCM brings about great improvements to the overall performance. Zhang et al.[251] showed that co-doping the TM and O sites with Al$^{3+}$ and (PO$_4$)$^{3-}$ respectively reduced cation mixing and improved the overall stability of the cathode and cycling ability, although with a slightly deleterious effect to the initial capacity.

Li and TM site co-doping with Na$^+$ and Al$^{3+}$ ions has been successfully demonstrated by Ko et al.[252] Na$^+$ ions with their relatively larger ionic radius expanded the c lattice parameter, while Al$_{3+}$ ions due to their lower electronegativity than Ni$^{3+}$ created more ionic M-O bonds, thus increasing the electrostatic repulsion between the O layers. The combined effects increased the c lattice parameter and improved lithium diffusion. Additionally, stronger Al-O bonds than Ni-O bonds reportedly stabilised the NCM layered structure and prevented lattice collapse.

Finally, simultaneous Li and O site doping has also been shown to enlarge the interlayer spacing, thus improving Li diffusion and rate capability[235,236].

*Challenges in doping*

Similar to wet-chemical coating procedures, NCM doping most often involves a calcination step at the end of material synthesis, which, analogously to the calcination step during coating, can cause migration of the introduced elements to the surface to form a coating. As has been shown for Zr$^{4+}$ dopant, temperatures as low as 400°C cause migration of Zr$^{4+}$ to the surface of the particles to form a Zr-containing coating[253]. Similarly, doping NCM with Sb was shown to also create an Sb oxides coating when annealing in O$_2$ atmosphere at 750 °C[254]. Thus, in cases where doping strategies are employed, a thorough study on the extent of coating should be performed by TEM, XPS, or other methods. However, most studies on doping disregard surface analysis and hence cannot exclude the presence of a coating and its effects on material properties[214,255].

Structural control during synthesis

It has been shown that altering the configuration of secondary particles decreases the structural and chemical degradation of NCM[256–259]. Several approaches have been proposed, each countering a different degradation phenomenon (figure 11). Radially ordering the primary particles within the secondary particle (figure 11b) reduces the detrimental effect of primary particle volume change, resulting in microcracking and secondary particle disintegration[256]. Using single-crystal NCM (figure 11a) also decreases the effects of particle volume change and microcracking; however, it notably impedes rate capability, as Li ions have fewer diffusion routes in a single crystal[260].

Another approach to control the structure of secondary particles is to create a core-shell particle (figure 11c), which functions similarly to a coating but involves a thicker layer that is usually chemically similar to the inner part and can be electrochemically active[261]. A core-shell particle could include a Ni-poor and Mn-rich outer layer (a shell) and a Ni-rich bulk (the core). The Ni-rich inner part would provide a high specific capacity, be protected from the effects of electrolytic degradation, and its volume change would be restricted due to a more rigid, stable outer layer. However, the core-shell interface may still degrade over time due to uneven expansion of the different materials. A smoother transition from a Ni-rich to Ni-poor region (a gradient) could alleviate such structural instabilities[262]. Such a gradient structure can also be obtained by doping the particle to various degrees as a function of the distance from the particle centre.

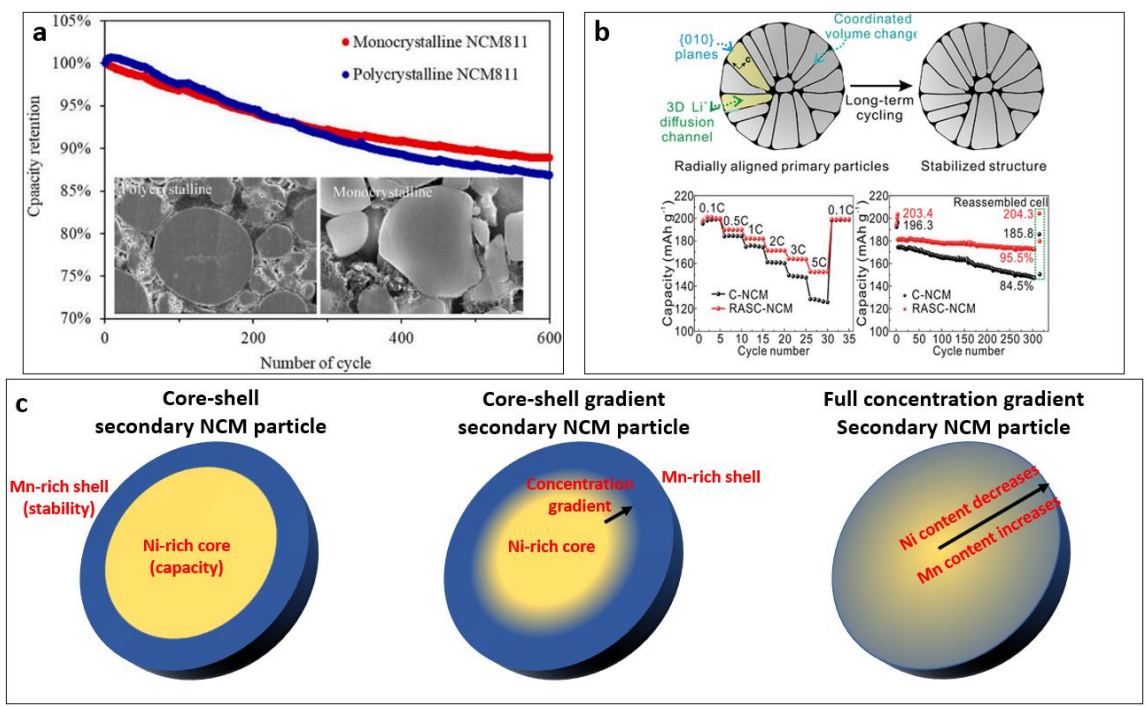

**Figure 11.** Types of structural modifications in NCM. a) Capacity retention of polycrystalline vs. monocrystalline NCM811, reprinted with permission from S. Niu et al.[263], copyright 2020 Elsevier, b) rate capability and capacity retention of commercial vs. radially aligned single crystal NCM811, reprinted with permission from X. Xu et al.[256], copyright 2019 WILEY-VCH Verlag GmbH & Co. KGaA, Weinheim, c) Illustration of core-shell, core-shell concentration gradient, and full concentration gradient NCM particle structures.

A continuous gradient structure (figure 11c) has been achieved by Sun et al.[262] by changing the Ni/Co/Mn ratio in the precursor solution of the co-precipitation synthesis by pumping Mn-rich solution into the starting Ni-rich solution during the reaction. The resulting particle structure contained a Ni-rich core and Mn-rich outer layer with a smooth transition between. The Co amount remained constant throughout the particle. When comparing to the inner composition and outer composition material cycling stability, the full concentration gradient (FCG) structure displayed capacity nearly as high as for the Ni-rich inner part, but with stability resembling that of the outer Mn-rich part. The full FCG material cell maintained a capacity retention of 90% even after 1000 charge-discharge cycles.

A gradient structure can also not be fully gradient from the core to the shell but integrate a transitional gradient layer between two compositions (core-shell gradient) which can alleviate the structural mismatch (figure 11c). As demonstrated by Li et al.[218], La-Al co-doping of NCM results in a complex structure, which displays a uniformly doped constant Ni-concentration (NCM811) bulk, transitional surface region (~1μm) where the Ni content gradually decreases and an outer layer of $La_2O_3$ which protects from surface-based reactions. Half cells containing this gradient cathode material displayed considerably improved capacity retention, retaining 80% of initial capacity after 400 cycles at 10C.

Considering the careful and tedious fabrication process, currently a gradient structure cathode material would be difficult to produce in large quantities for use in EVs or other mobile energy storage applications. However, as gradient structures exhibit such promising characteristics, the research in this field continues[264,265].

### 4.2. Degradation mitigation strategies in NCMs

Investigation of the currently available research has revealed an extensive collection of electrochemical results obtained over a wide range of parameters. Unfortunately, the sheer amount and variance of parameters (e.g., particle size and morphology, electrode formulation, electrolyte, voltage range and corresponding SoC, temperature, current density used for cycling, etc.) make a direct comparison between any single studies ineffective. This hinders the formation of any definite conclusions unless a systematic single, all-encompassing study is performed.

However, when reviewing results jointly, certain trends appear and can be analysed without a direct focus on the peculiarities of any single study. Although data is not abundant, many tendencies in the effects of coating vs. doping can be seen as a function of Ni-content in NCM electrodes. Figure 12 comprises the results from 45 publications describing the extent to which doping, or surface coating has improved the cycle life of pristine NCM111, NCM622, and NCM811. Due to many researchers limiting their cycle life studies to 50 cycles and many more to only 100 cycles, two graphs show capacity retention at 50 and 100 cycles (Figure 12). Improvement of capacity retention (%CR) shown on the vertical axis is calculated in percentage points as the additional capacity obtained in the 50[th] (or 100[th]) cycle of the treated material (CR$_{treated\ NCM}$) when compared to the reference material used in the study (CR$_{reference\ NCM}$).

$$\%CR = \frac{CR_{treated\ NCM} - CR_{reference\ NCM}}{CR_{reference\ NCM}} \qquad \text{(Eq. 1)}$$

Relative units are compared instead of absolute cycle life values due to widely differing cycle lives of reference (pristine) materials. Of course, the comparison of such a dataset is inherently complex. If the reader wants to draw their conclusions, the data table is provided in the supplementary information.

For NCM111, the conclusion is perhaps the most straightforward one – statistically, coating yields larger improvements in cycle life than doping. In NCM111, the SoC (or degree of delithiation) at the same operating voltage range is much lower and lattice parameters do not change as much as for high-Ni NCMs, thus doping strategies to mitigate the little to no microcracking in NCM111 are not effective for cycle life extension. Instead, surface protection provides better mitigation of degradation in NCM111 due to a larger SoC being reached, and thus more rapid degradation occurring on the surface than in bulk. For NCM622, the evolution of SoC with voltage is similar to NCM111[40]; however, critical SoC where lattice parameter change initiates is reached at a slightly lower voltage. Thus, a combination of doping and coating most effectively improves the cycle life in NCM622. As NCM811 experiences more significant lattice parameter change at lower voltages due to a higher SoC being reached, doping strategy to mitigate degradation in NCM811 becomes more beneficial. However, coating to mitigate surface and electrolyte degradation should also be employed for the most effective cycle life extension of NCM811. Mitigation strategies for each NCM material are viewed in greater detail in the subsequent chapters.

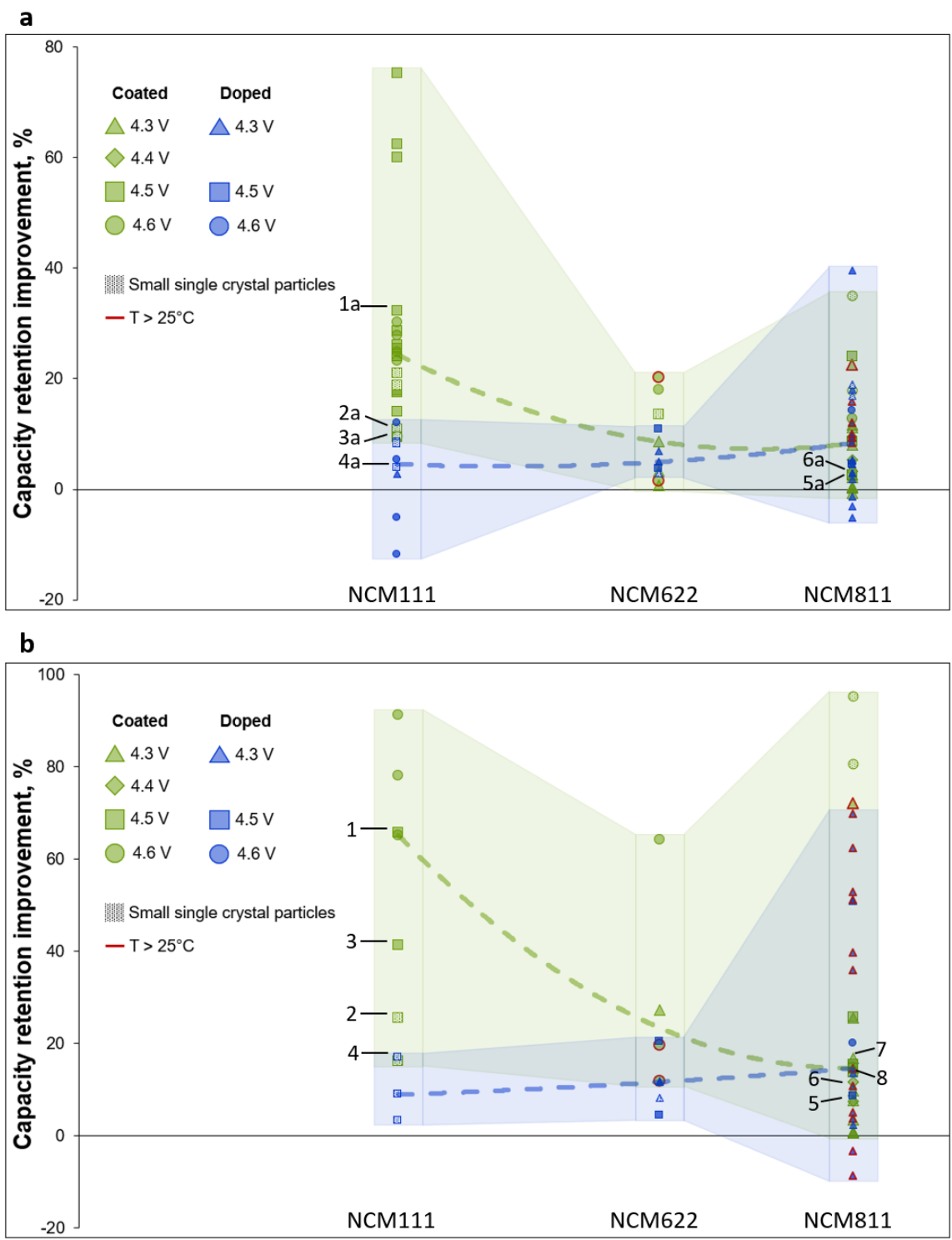

**Figure 12.** Capacity retention improvement (calculated as shown in equation 1) after a) 50 cycles and b) 100 cycles of coated and doped NCM cathodes compared to the reference untreated samples as reported in 45 literature sources (for reference data, see Supplementary information).

NCM111

NCM111 cathode material is structurally stable at potentials up to 4.3 V vs. Li/Li$^+$ due to the large amount of Mn stabilizing the structure and lower SoC being reached, showing little to no microcracking during cycling, as discussed in sections 2.3. and 3.1. It has been commercially used without any modifications and exhibits decent capacity retention[49]. However, the cycling stability and rate capability could be further improved by mitigating the detrimental effects of RLCs left on the surface after synthesis, hindering surface phase transitions and halting CEI growth.

Since most degradation of NCM111 occurs on the surface of the material, based solely on the generally proposed effects of coating (surface protection) and doping (bulk lattice stabilization), the cycling stability would most effectively be improved by coating the NCM111 material. It is confirmed by the compiled data in figure 12, which shows that over all voltage ranges and particle sizes both at 50 cycles and at 100 cycles the most improvement in capacity retention is achieved by surface coating of NCM111 particles.

Surface coating is the most widely studied method for the improvement of surface stability in NCM111 cathode materials (see supplementary information)[158,266–270]. Qiu et al.[158] showed that coating NCM111 particles with $Al_2O_3$ significantly improves battery performance when cycling up to 4.5V. Capacity retention in 5% coated NCM111 samples was as high as 97.5% after 100 cycles. In contrast, the bare NCM111 material showed capacity retention of 58.8%, with %CR of 65.8% (point 1 in figure 12b) or 32.3% after 50 cycles (point 1a in figure 12a). Similar results have been shown by Chen et al.[266] (%CR=25.6%, point 2 in figure 12b or %CR=10.8%, point 2a in figure 12a) and Liu et al.[268] (%CR=41.5%, point 3 in figure 12b or %CR=9.4%, point 3a in figure 12a), independent of the coating material. Whereas the doping effect for NCM111, also cycled up to 4.5 V for 100 cycles, has been shown by C.X. Ding et al.[271] (point 4 in figure 12b; point 4a in figure 12a for 50 cycles). In this case, the capacity retention increased from 79.2% for the pristine material to 92.7% for the Zr-doped

NCM111 (%CR=17.1%). Furthermore, no comment was made regarding the effect of the sintering step (950°C, 15h) on the migration of Zr towards the surface of the particle to form a Zr-containing coating, which means that the increase in capacity retention of the doped NCM111 could partly be due to protective effects of a Zr-containing coating, as discussed above.

The same trend can be observed, when the compiled electrochemical data is converted to capacity retention improvement after 50 cycles. In general, doped samples show lower capacity retention improvement than coated samples, indicating the superiority of coating for cycle life improvement of NCM111.

Overall, the electrochemical data gathered for 100 and 50 cycles confirms that surface coating is the best strategy for mitigating the surface-related degradation effects of NCM111 cathode material.

### NCM622

Since NCM622 experiences comparatively larger lattice parameter changes during cycling compared to NCM111 (figure 3), which was shown to cause microcracking when cycling up to 4.2 V[142], doping the material to stabilize the structure is more beneficial than doping NCM111[49]. When compared to NCM811, on the other hand, the lattice parameter changes and hence microcracking are not as severe[141]. Based on these considerations, surface protection looks to be most beneficial for mitigating the degradation in NCM622. However, introducing doping could mitigate the degradation of NCM622 associated with microcracking and bulk degradation.

When looking at the research of coated and doped NCM622, similarly to NCM111, most authors do not analyse the migration of coating material within the bulk to dope the surface layer or discuss the diffusion of the doped species to the surface of the cathode particle to form a coating. Nevertheless, the electrochemical data for coated and doped NCM622 indicate that coating still improves the cycling stability to a greater extent than doping. However, doping effects in the case of NCM622 are not insignificant and improve the capacity retention of pristine NCM622 as much as some of the less enhanced coated samples.

Thus, the analysis indicates that the optimal mitigation strategy for improving NCM622 cycle life combines surface protection and bulk stabilization with a focus on surface coating. Perhaps introducing a coating material which diffuses within the particle to dope and stabilize the structure could be considered.

### NCM811

An effective strategy to mitigate the detrimental effects of the *c* lattice parameter contraction could significantly improve the structural stability of NCM811 cathode material. Many such attempts have been made by doping the cathode material and successfully enhancing battery lifespan[207,216,219,224,225,272]. As surface degradation also occurs to a larger extent in NCM811 than in NCM622 and NCM111, coating the cathode particles should also be employed to successfully mitigate the degradation.

Based on the evaluated research compiled in figure 12, surface and bulk degradation play equally significant parts in the degradation of NCM811 cathodes. If considered separately, on average, both doping and coating of NCM811 yield similar improvements in cycle life. Thus, based on the statistical survey results, it appears that the most appropriate strategies to mitigate the degradation of NCM811 would call for both – surface and bulk stabilization.

The capacity retention of coated vs. doped NCM811 was evaluated by comparing research by Hu et al.[273] and Jiang et al.[219] (points 5 and 6 in figure 12b and points 5a and 6a in figure 12a). The cathode material was coated with $Al_2O_3$ or doped with $Ti^{4+}$, respectively. The NCM811 cathode materials modified within these two publications were cycled within the voltage range of 2.8-4.3 V, with a current density of 1C (200 mA/g) for 200 cycles at room temperature. $Al_2O_3$-coated cathode material displays very similar capacity retention after 200 cycles (95.4%) to the Ti-doped material (95.0%). The %CR after 100 cycles is 7.8% and 10.8%, and after 50 cycles – 2.1% and 3.0%, respectively. If no coating material and dopant diffusion is assumed, both modifications are equally important in mitigating the degradation of NCM811. Similar results were also obtained by Huang et al.[274] ($MoO_3$-coated NCM811 showed capacity retention of 94.8% after 100 cycles to 4.3 V at 1 C; %CR=14.8%) and Li et. al.[225] (Mg-doped NCM811 showed capacity retention of 91.9% after 100 cycles to 4.3 V at 1 C; %CR=13.6%), corresponding to points 7 and 8 in figure 12b. Comparing coating and doping effects on cycle stability in figure 12 overall, the average improvement in capacity retention compared to the pristine material is comparable for coated and doped NCM811. It should be noted, however, that in some cases, the "pristine" material, if ordered for experimental needs from commercial sources, may be pre-doped or pre-coated by the manufacturer, making the baseline data somewhat misleading if a proper analysis of the starting material is not carried out.

It has been shown that doping NCM with Al reduces c lattice parameter change during charge from 3.2% to 2.7%. However, it does not fully remove the lattice collapse occurring around 80% SoC[275]. Therefore, more advanced doping strategies or microstructure modifications should be employed to eradicate microcracking completely. As in the previously described case of $LiNi_{0.8}Mn_{0.13}Ti_{0.02}Mg_{0.02}Nb_{0.01}Mo_{0.02}O_2$ developed by Zhang et al.[72], the journey of eliminating volume or lattice parameter changes can lead to the development of materials that can deviate from the typical NCM811 stoichiometry.

# 5. A GLIMPSE BEYOND NCM811

## 5.1. Ni-Rich NCMs – NCM83 and NCM90

State-of-art Li-ion battery cathode materials are currently advancing to compositions beyond NCM811, further increasing the Ni content. This enables obtaining larger capacity and both specific energy and energy density. Increasing the Ni content to 83% (NCM83) and 90% (NCM90) comes with even more aggravated surface and bulk degradation, creating further challenges and opportunities for material scientists. The RLCs on the surface of the particles become even more abundant, cation mixing occurs more easily due to a larger number of $Ni^{2+}$ ions with similar ionic radii as $Li^+$, the related surface reconstruction and oxygen evolution become more rapid, the lattice parameter changes are even more extreme than for NCM811, causing microcrack formation and particle disintegration and the peak in the dQ/dV plot (figure 5), widely associated with rapid degradation in NCMs is even sharper and more pronounced[43,49,92,141,260,276]. All in all, as shown in figure 1b, while the capacity grows with the content of Ni, increasing efforts should be devoted to improving the cycle life and stability of such cathodes.

An important degradation phenomenon that should be explored for $LiNi_xMn_yCo_zO_2$ compositions with 0.8<x<1 is the H2 → H3 phase transition which occurs in $LiNiO_2$ but has been rebutted in NCM811 by Clare P. Gray's group[37]. The discovery of the composition at which the phase transition starts occurring would provide the upper threshold value of Ni content, and hence, the highest capacity, which can be achieved in cathodes, simultaneously avoiding the detrimental phase transition. From there on, the longevity and cyclability of NCM with the threshold composition could be improved by doping, coating, and optimizing the microstructure.

Overall, the discussed trends in degradation with increasing Ni content continue in NCM83 and NCM90. Hence, the trend in the most effective mitigation strategies depicted in figure 12 may also be extended to NCM83 and NCM90. As the nickel content increases beyond 80%, the effect of lattice parameter changes and microcracking becomes even more pronounced[141,260,276], suggesting doping as the most effective degradation mitigation strategy, surpassing coating if only one of the strategies is to be chosen. This follows the trend outlined in figure 12, where doping gradually becomes the more valuable materials development technique with increasing Ni content. A detailed data collection and analysis similar to that assembled in the supplementary information and depicted in figure 12 could be conducted to verify this assumption once a sufficient amount of data becomes available.

A recent study explores the effects of Mg doping and $Li_3PO_4$ coating on NCM83[277]. It shows that implementing both strategies gives the best improvement in capacity

retention – 90% capacity retention after 200 cycles vs. 70.1% capacity retention of the reference. Comparing the individual effects of doping and coating it shows that doping gives rise to better capacity retention (84.5%) than coating (80.5%). Hence, this study hints towards the trend in doping effects indeed surpassing that of coating; however, if a reliable conclusion is to be drawn, substantiating doping and, hence, microcracking effects over surface degradation/protection, significantly more evidence should be gathered.

In the case of NCM90, the same trend is expected to continue, showing that doping improves the cathode performance to a greater extent than coating. However, in all Ni-rich (Ni>80%) NCMs, a combination of coating and doping strategies is presumably the best approach to mitigate cathode degradation[92,141,179].

### 5.2. Beyond NCM

Most of the degradation pathways and mitigation strategies discussed in this paper are not exclusive to NCM materials. Crack formation, structural changes in the crystal lattice and reactions with electrolyte can, to a varying degree, be attributed to phosphates, spinel-type materials and other electrode materials. Similarly, coating, doping, and microstructure tailoring serve as degradation mitigation strategies for many other cathode materials, including $LiFePO_4$, other layered transition metal oxides, spinel-type materials[19,278–281], and even materials for lithium – chalcogen (especially lithium – sulphur[282]) battery cells or even metallic lithium anodes[283].

To illustrate a few examples, NCA or $LiNi_xCo_yAl_zO_2$ (x+y+z=1 in full), one of the alternatives to NCM layered transition metal oxides suffers from similar ageing issues, including reactions with electrolyte at high electrode potentials and subsequent surface phase transitions/reconstruction (spinel and rock-salt structures) and gas evolution[43,284], and microcracking[146]. Hence, the set of mitigative approaches include coating the surface of the electrodes, introducing dopants, and tailoring the microstructure of the electrode. High voltage spinel $LiMn_{1.5}Ni_{0.5}O_4$ can operate at higher voltages of 4.7 V, and is more stable if TM sites are doped (this eliminates Jahn-Teller distortion, hinders the formation of the rock salt impurity phase, and stabilizes the disordered structure[285]) and if particles are coated (this allows operation at high voltages mitigating LLI and decomposition of electrolyte[172,286,287]).

Finally, the degradation pathways and their mitigation methodology can be applied for battery cells based on other mobile ionic species. Most notably, layered P2-type or O3-type $Na_{2/3}MO_2$ (M – transition metal) materials are among the highest capacity cathodes for sodium-ion batteries[288]. Like NCM materials, $Na_{2/3}MO_2$ materials also suffer from undesired transformations of the crystal lattice[289–291], irreversible anion redox reactions resulting in oxygen evolution[74,292,293], excessive CEI growth[294–296], microcrack formation[297] and dissolution of manganese in the low potential range[298]. It follows that the strategies to mitigate the degradation of $Na_{2/3}MO_2$ cathodes are also analogous. Doping, especially in the TM site, has been a widely studied and useful approach[299,300], as well as coating, both with carbon and electric insulators[161,301–303], and tailoring of the electrode microstructure[304].

### 5.3. Full battery cell

In a full battery cell, degradation of the cathode is not an isolated phenomenon and impacts the whole environment of the cell, and vice versa (brief discussion in part 2.4.). Hence, reduced cathode degradation is also often a consequence of anode or electrolyte improvements or modification.

Functional additives have often been introduced into the electrolyte to stabilise the growth of CEI or SEI, or to remove reactive species (HF, $H_2O$) from the electrolyte[305]. For example, Pham et al.[306] showed that adding 1% vinylene carbonate (VC) to non-flammable $LiPF_6$/PC/FEMC/DFDEC electrolyte is effective in stabilizing the CEI on NCM811, in addition to reducing crack formation, TM dissolution, and structural degradation, even when charged up to 4.5 V. As a result, the cycling stability and capacity retention was improved from 76% to 82% after 50 cycles. Another report by Lee et al.[307] shows that adding fluoroethylene carbonate (FEC) and 1,1,2,2-tetrafluoroethyl-2,2,3,3-tetrafluoropropyl-ether (TTE) to the electrolyte stabilizes not only the CEI, but also the SEI by forming a LiF coating. Furthermore, they showed that if the SEI on a Li anode is damaged during cycling, TTE can repair the SEI by migrating to the exposed Li anode parts due to the F site on TTE being lithiophilic. Consequentially, the cycling stability was greatly improved.

Mitigating the degradation on the anode side usually involves suppressing Li dendrite growth, stabilizing the SEI, and creating a structurally stable host for lithium[308–310]. Stabilizing the SEI is often achieved by utilising electrolyte additives, and their effects are usually not isolated to the anode[310]. In such cases mitigating anode degradation can have a positive impact on the cathode side as well. Preventing Li dendrite growth means that LLI is reduced and more lithium can participate in cell operation. Thus, it might not affect the cathode degradation directly, however, the capacity of the whole cell is maintained and cycling stability improved.

## 6. SUMMARY AND OUTLOOK

NCM materials originate from layered oxide cathode materials. They have evolved towards higher nickel content ever since, as it allows increasing energy density, lowering the operating voltage to reasonable levels, and alleviating some of the sustainability-related concerns of the material. However, with increasing nickel content, many more stability-related issues start to appear in NCM materials. They range from decreased stability in ambient atmosphere to particle microcracking, undesired phase transitions, oxygen evolution, cathode-electrolyte interphase growth, transition metal dissolution, and degradation of solid-electrolyte interphase. A critical review of the literature suggests that the dominant mechanisms of ageing differ depending on the Ni content, so the most effective strategies for mitigating the decomposition of NCM should also be viewed in the context of Ni stoichiometry.

Summarizing the effects of coating and doping on degradation mitigation of NCMs with different Ni-content (table 1), it becomes apparent that with increasing Ni content and lowering of the voltage at which a critical SoC is achieved, the importance of stabilizing the crystal lattice and mitigating its volume change by doping increases. In NCM111 where the main degradation was shown to be surface-related, coating is significantly more effective in performance improvement than doping. In NCM622, microcracks start playing an increasingly important role in cathode degradation and, thus, doping to stabilize the structure of NCM becomes comparatively more efficient in degradation mitigation. In NCM811, the structural instabilities are more pronounced, thus the effect of doping as a mitigation strategy becomes just as important as that of coating. For compositions exceeding 80% Ni, both doping to alleviate the more extensive lattice parameter change when charged to exceeding lower Li content and coating to stabilize the more rapid surface-based decomposition reactions should be strongly considered due to the increasing rate of degradation and the multitude of degradation pathways[277].

**Table 1**. Summary of degradation mechanisms for various NCM materials and corresponding mitigation strategies.

| Composition | Degradation | Mitigation |
| --- | --- | --- |
| NCM111 | Cation mixing, surface reconstruction, oxygen release, RLCs, CEI formation | On average, coating strategy is superior to doping |
| NCM622 | Cation mixing, surface reconstruction, oxygen release, RLCs, CEI formation, microcracking | Coating provides more significant improvement; doping strategies also notably improve performance |
| NCM811 | Cation mixing, surface reconstruction, oxygen release, RLCs, CEI formation, significant microcracking | Coating and doping strategies provide comparable improvement to capacity retention |

As the trend towards higher voltage and higher capacity battery cells continues, the toolset used for analysing, categorising, and mitigating the stability-related issues of cathode materials can be relevant to many other cathode chemistries, including Li-rich NCM, high voltage spinel, and even cathodes for Li – sulphur batteries. With layered oxide materials playing an important role in post-lithium batteries and Na-ion batteries in particular, the knowledge of the decomposition of layered transition metal oxide materials and the set of strategies mitigating these can also have far-reaching consequences beyond Li-ion battery chemistry.

## Acknowledgements

Authors thank Peter Axmann for his valuable advice during the preparation of the manuscript. This research was funded by the Latvian Council of Science project "Cycle life prediction of lithium-ion battery electrodes and cells, utilizing current-voltage response measurements", project No. LZP-2020/1–0425. Institute of Solid-State Physics, University of Latvia, as the Centre of Excellence, has received funding from the European Union's Horizon 2020 Framework Program H2020-WIDESPREAD-01-2016-2017-TeamingPhase2 under grant agreement No. 739508, project CAMART2.

## Author contributions

LB was responsible for conceptualization, data curation, formal analysis, methodology, visualisation, writing of original draft, review & editing; MM: writing - review & editing; GK: conceptualization, funding acquisition, methodology, project administration, resources, supervision, validation, writing of original draft, review & editing.

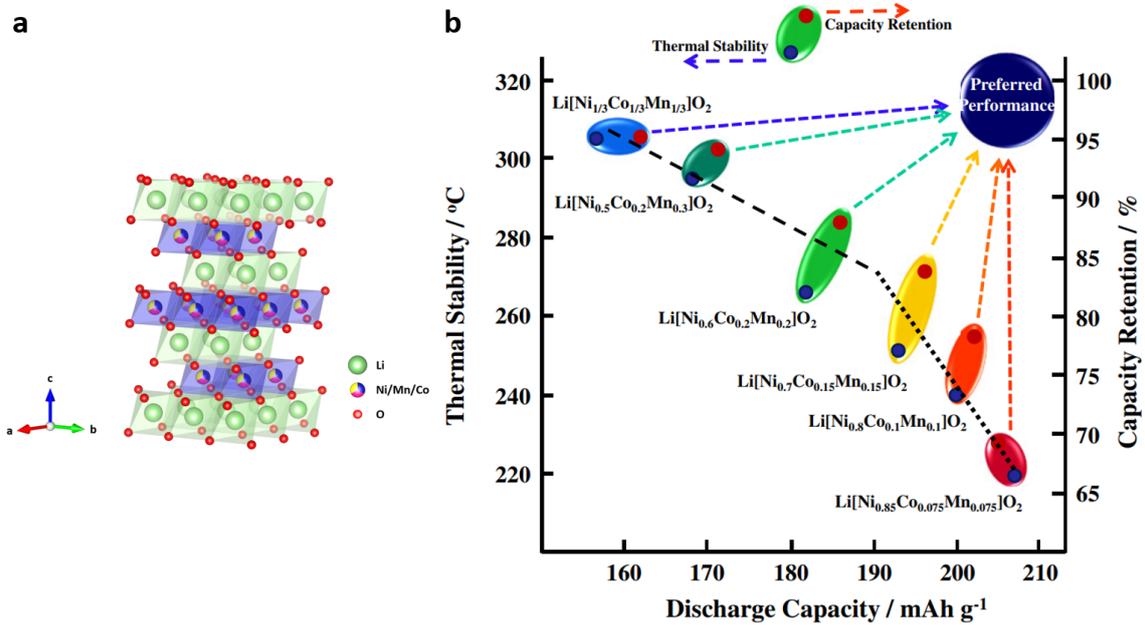

**Figure 1.** a) Crystallographic structure of NCM, b) Discharge capacity, capacity retention, and thermal stability of NCMs with different Ni composition. Reprinted with permission from H.-J. Noh et al.[49], copyright 2013 Elsevier.

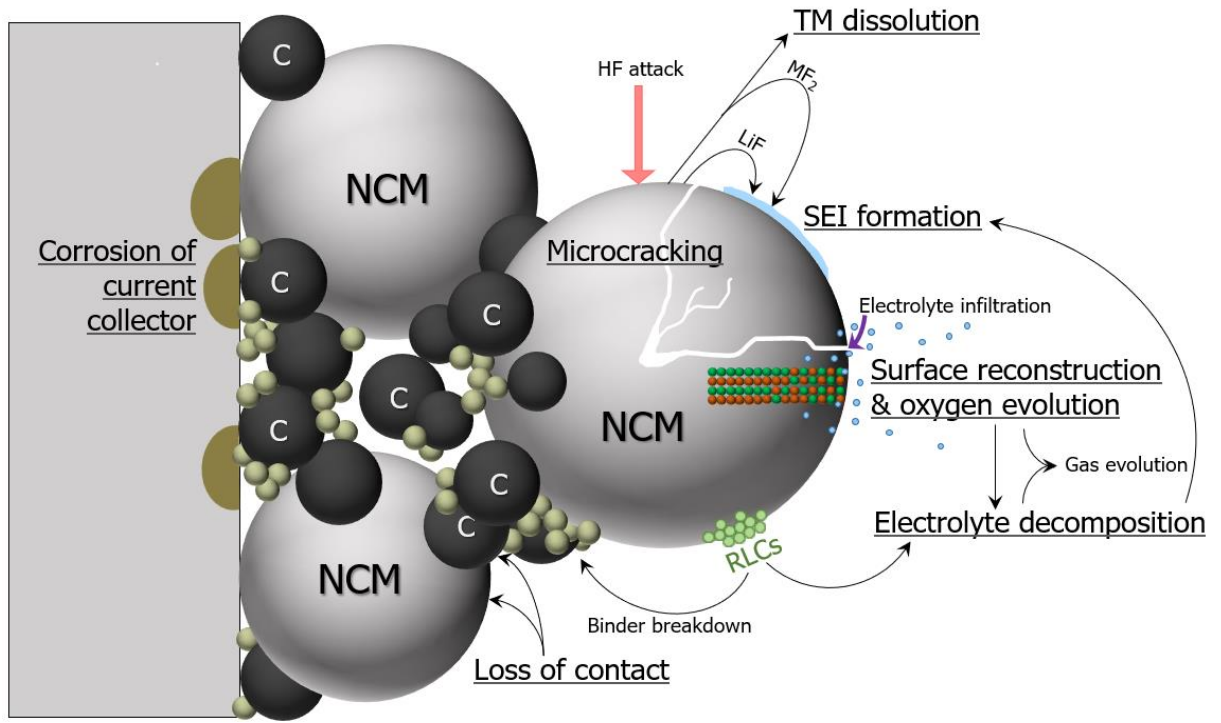

**Figure 2.** Ageing mechanisms of NCM cathode materials.

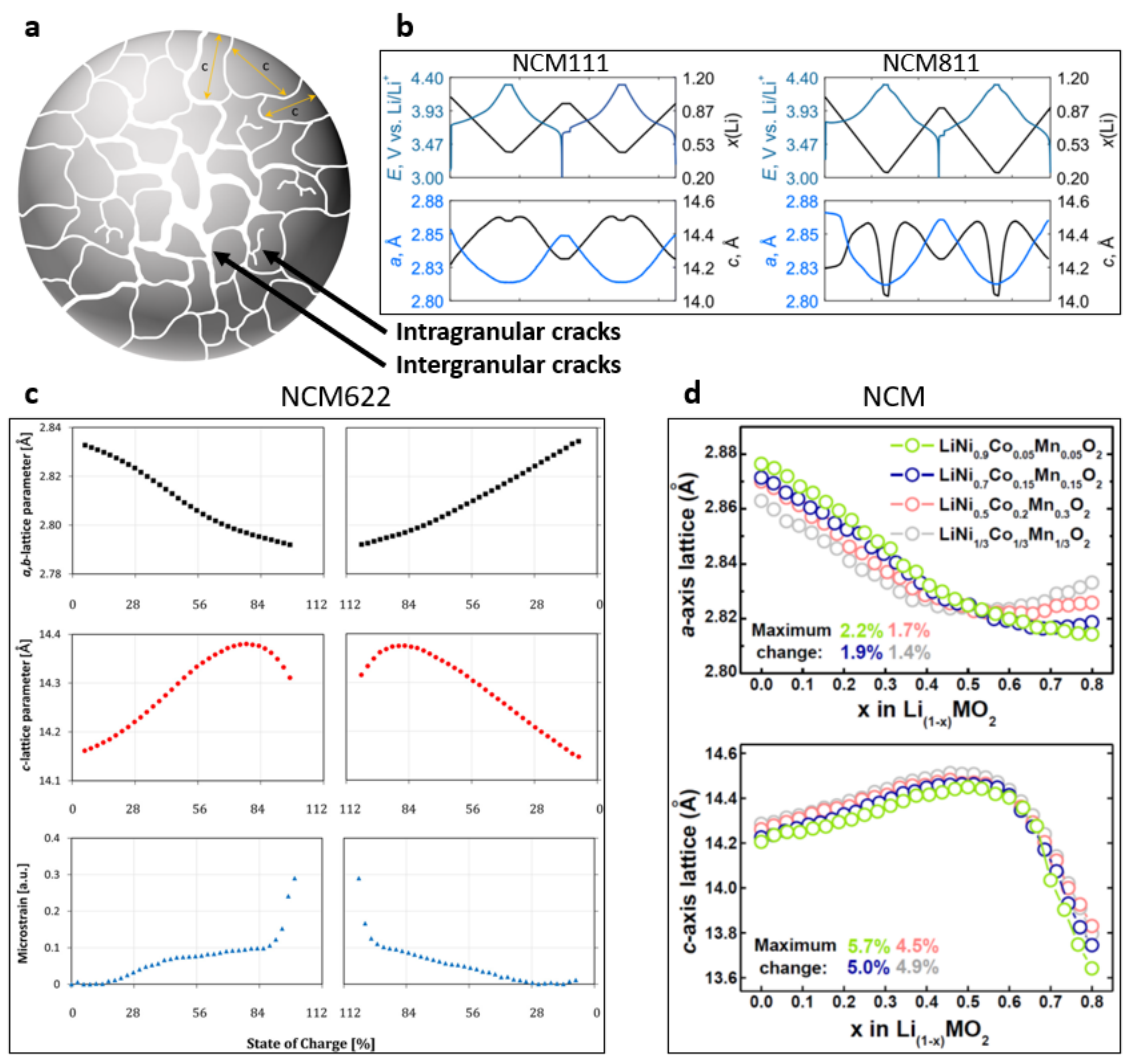

**Figure 3.** a) Illustration of intergranular and intragranular cracks and anisotropic lattice orientation in NCMs, b) changes in lattice parameters during charge-discharge in NCM111 and NCM811, reprinted with permissions from A.O.Kondrakov et al.[69], copyright 2017 American Chemical Society, c) changes in lattice parameters and microstrain during charge-discharge in NCM622, reprinted from ref.[73], licenced under CC BY-NC-ND 4.0, d) changes in lattice parameters during charge-discharge in NCMs with different stoichiometry, reprinted with permission from L. Wangda et al.[39], copyright 2019 American Chemical Society.

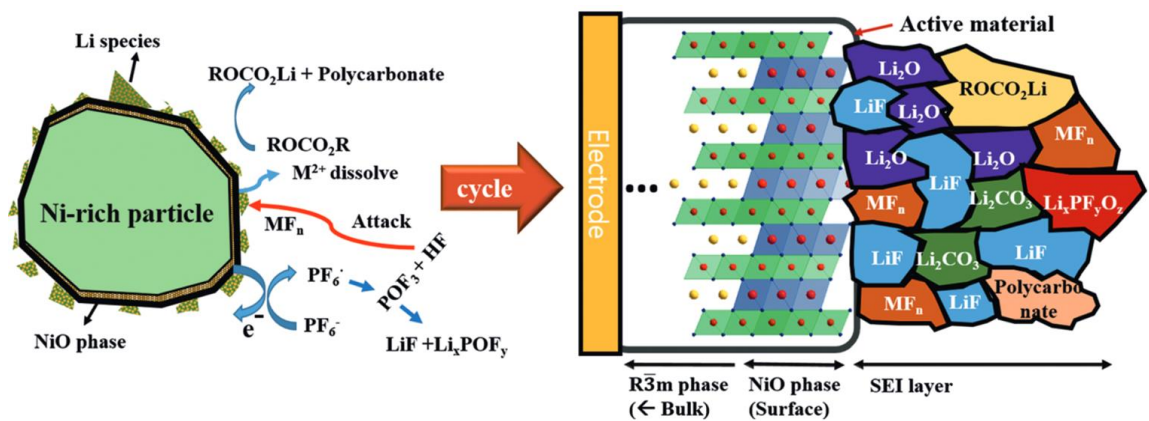

**Figure 4.** Evolution and composition of CEI, reprinted with permission from W. Liu et al.[32], copyright 2015 WILEY-VCH Verlag GmbH & Co. KGaA, Weinheim.

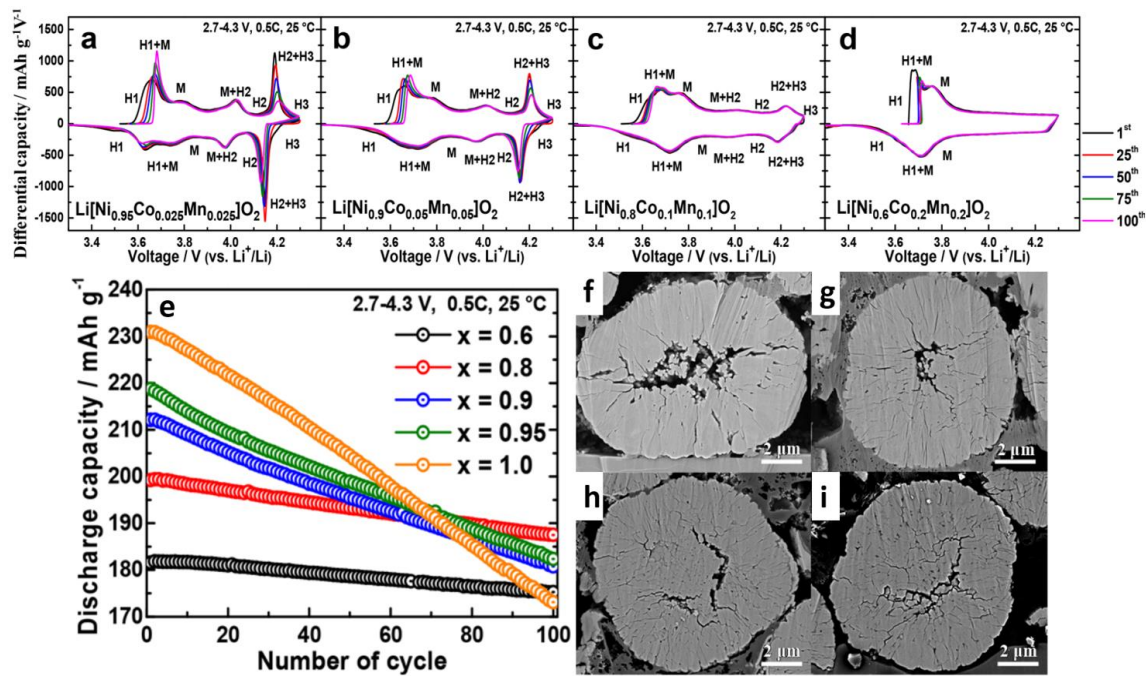

**Figure 5.** dQ/dV plots of a) NCM96, b) NCM90, c) NCM811, and d) NCM622; e) cycling curves of the respective composition cells; f-i) Cross-sectional SEM images of f) NCM622, g) NCM811, h) NCM90, and i) NCM96 cathode materials in the fully charged state (at 4.3 V in the first charge), Reprinted with permission from H.-H. Ryu et al.[141], copyright 2018 American Chemical Society.

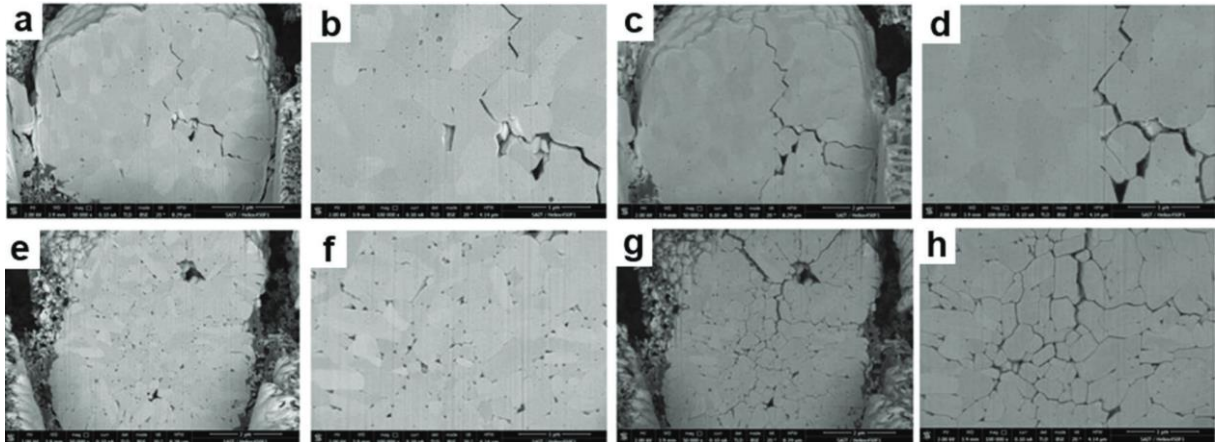

**Figure 6.** Microcrack distribution in NCM111 (a-d) and NCM811 (e-h) before (a-b, e-f) and after (c-d, g-h) being charged to 4.35 V for 50 cycles, reprinted from ref.[136], licenced under CC BY 4.0.

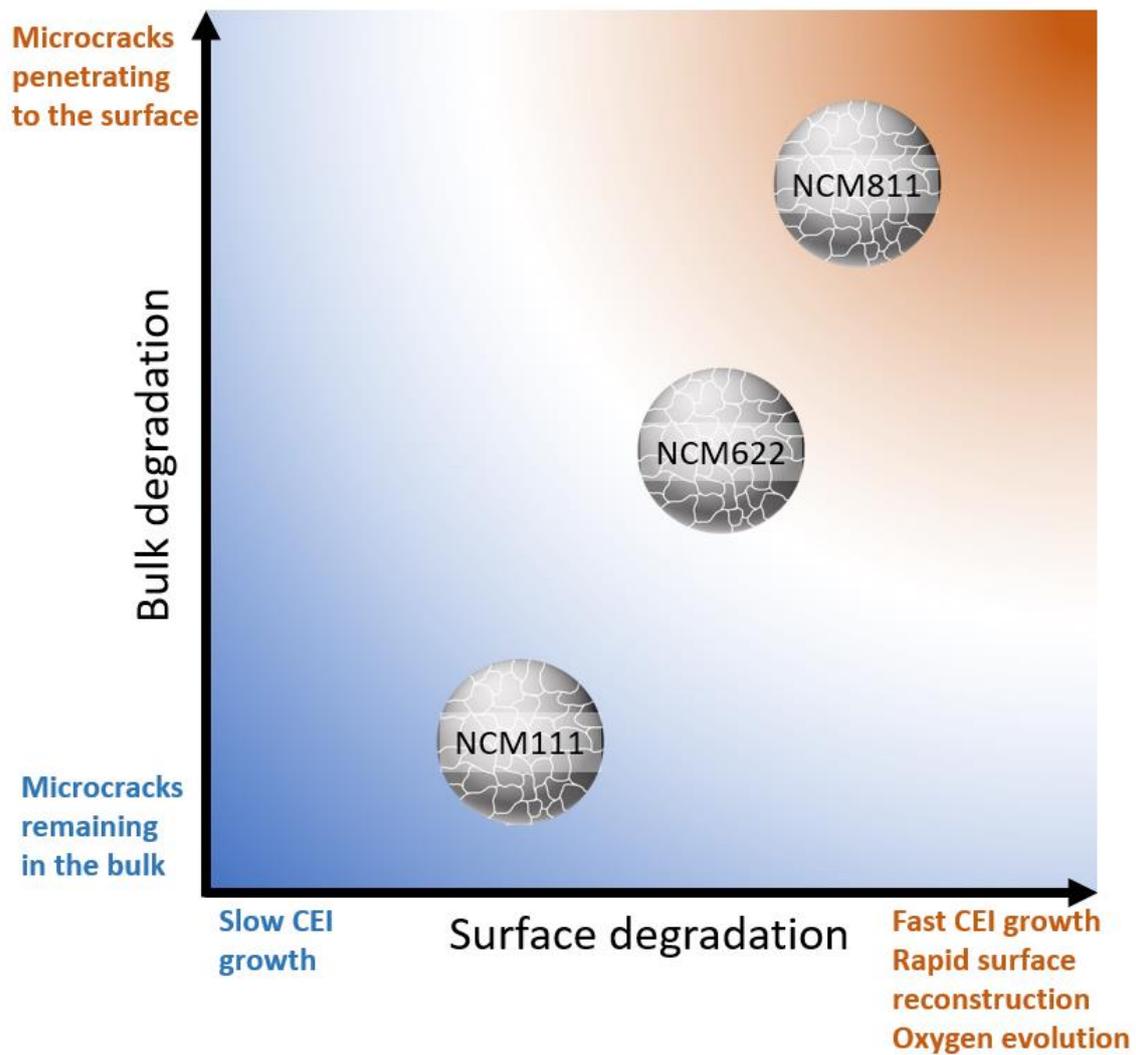

**Figure 7.** Illustration of the main type of degradation occurring in the NCM111, NCM622, and NCM811 cathodes during battery operation.

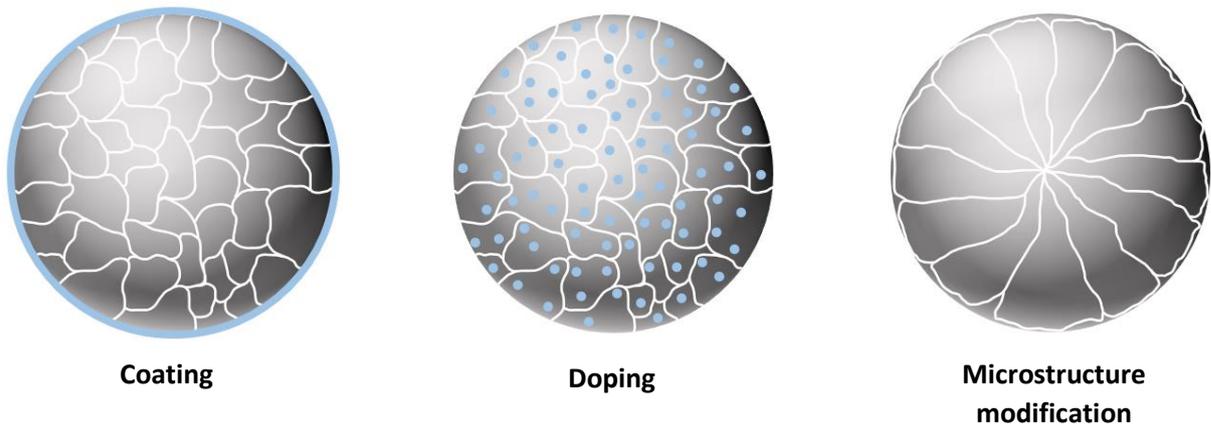

**Figure 8.** Illustration of the main cathode degradation mitigation strategies – coating, doping, and microstructure modification.

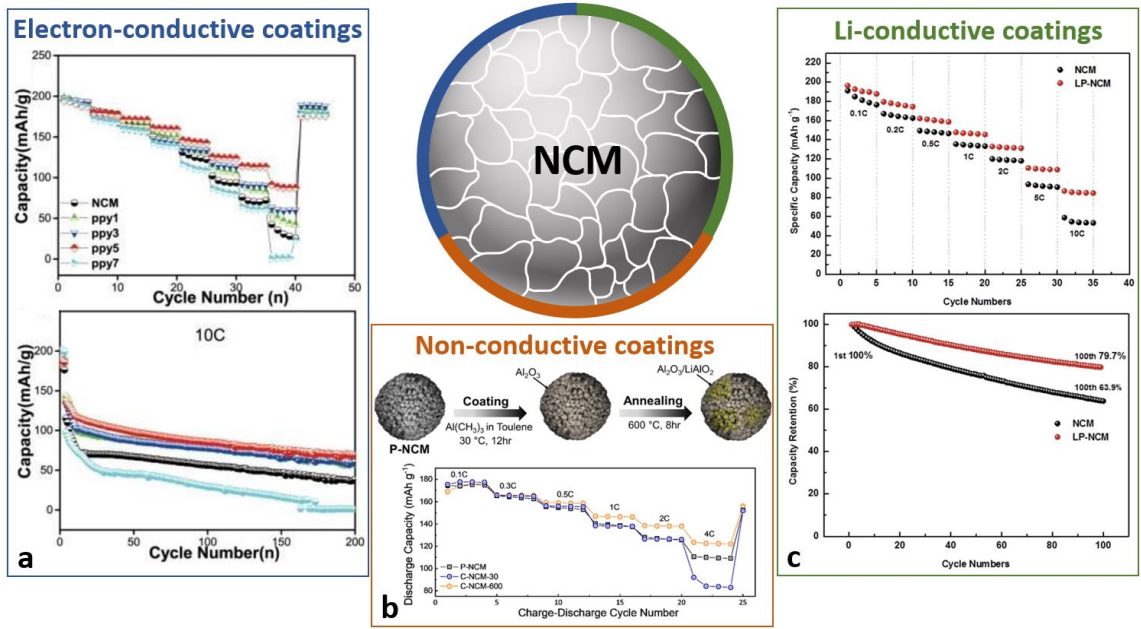

**Figure 9.** Types of coatings on NCM and their effect on the electrochemical properties. a) Effect of polypyrrole coating on the rate capability and capacity retention of NCM811, reprinted with permission from B. Li et al.[171], copyright 2019 Wiley-VCH Verlag GmbH & Co. KGaA, Weinheim, b) effect of $Al_2O_3$ and $Al_2O_3/LiAlO_2$ coating on the rate capability of NCM701515, reprinted with permission from R. S. Negi et al.[175], copyright 2021 American Chemical Society, c) effect of a $Li_3PO_4$ coating on the rate capability and capacity retention of NCM622, reprinted with permission from S.-W. Lee et al.[176], copyright 2017 Elsevier.

**Figure 10.** Doping sites in NCM and their effect on the electrochemical properties. a) Effect of F doping in O sites on the capacity retention of NCM811, reprinted from ref.[229], licenced under CC BY, b) effect of Al doping in TM sites on the capacity retention of NCM761014, reprinted with permission from W. Zhao et al.[237], copyright 2020 American Chemical Society, c) effect of Na doping in Li sites on the capacity retention and rate capability of NCM600535, reprinted with permission from Y. Shen et al.[238], copyright 2021 Elsevier.

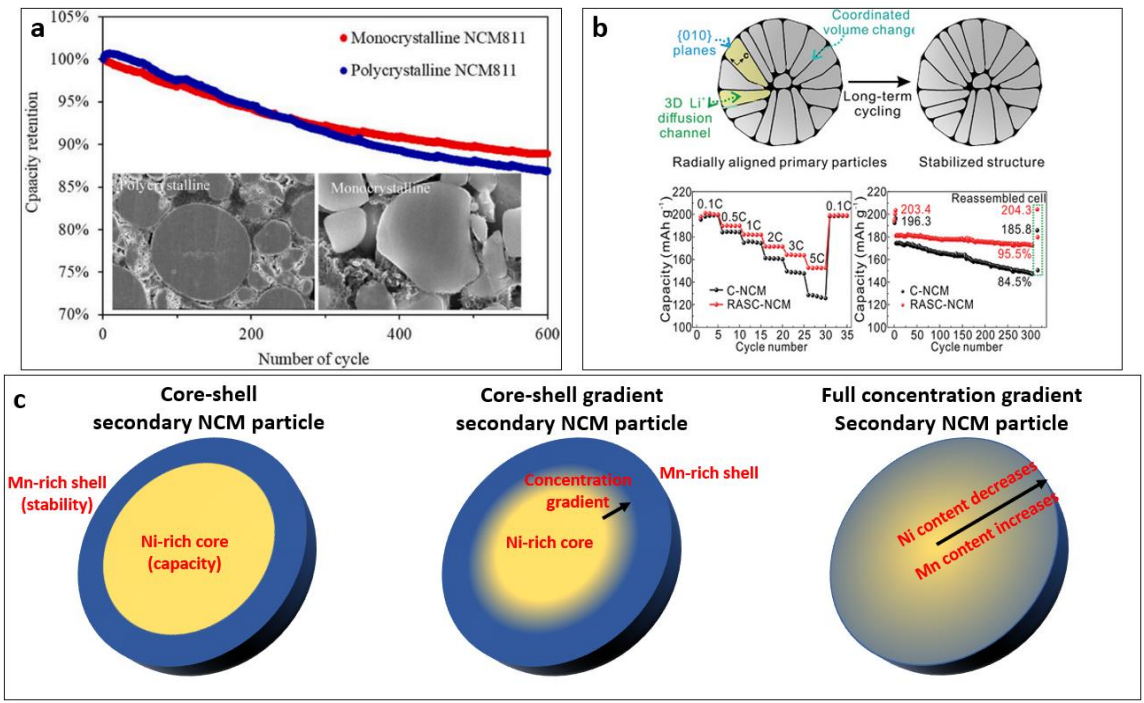

**Figure 11.** Types of structural modifications in NCM. a) Capacity retention of polycrystalline vs. monocrystalline NCM811, reprinted with permission from S. Niu et al.[263], copyright 2020 Elsevier, b) rate capability and capacity retention of commercial vs. radially aligned single crystal NCM811, reprinted with permission from X. Xu et al.[256], copyright 2019 WILEY-VCH Verlag GmbH & Co. KGaA, Weinheim, c) Illustration of core-shell, core-shell concentration gradient, and full concentration gradient NCM particle structures.

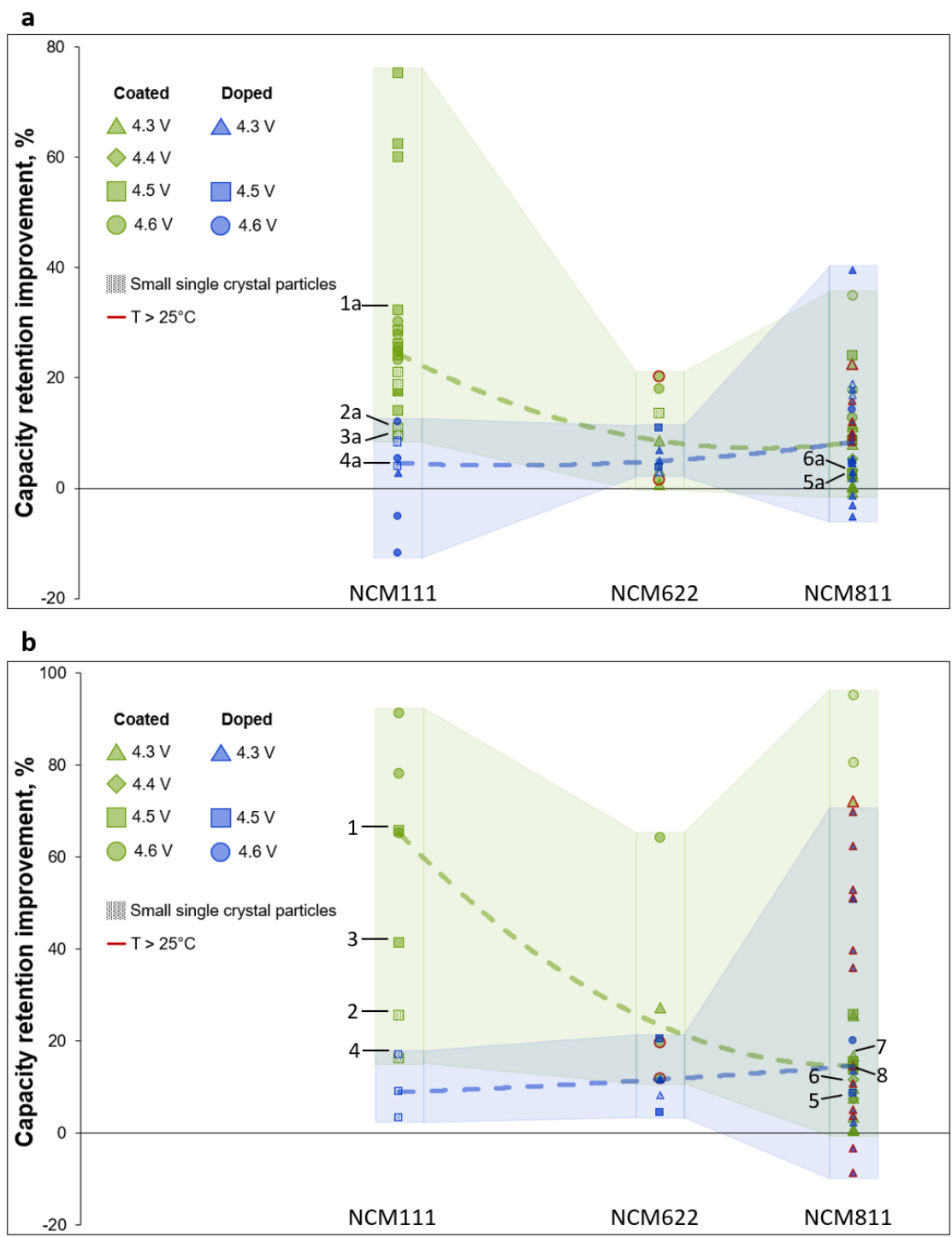

**Figure 12.** Capacity retention improvement (calculated as shown in equation 1) after a) 50 cycles and b) 100 cycles of coated and doped NCM cathodes compared to the reference untreated samples as reported in 45 literature sources (for reference data, see Supplementary information).

## Tables

**Table 1**. Summary of degradation mechanisms for various NCM materials and corresponding mitigation mechanisms.

| Composition | Degradation | Mitigation |
|---|---|---|
| NCM111 | Cation mixing, surface reconstruction, oxygen release, RLCs, CEI formation | On average, coating strategy is superior to doping |
| NCM622 | Cation mixing, surface reconstruction, oxygen release, RLCs, CEI formation, microcracking | Coating provides more significant improvement; doping strategies also notably improve performance |
| NCM811 | Cation mixing, surface reconstruction, oxygen release, RLCs, CEI formation, significant microcracking | Coating and doping strategies provide comparable improvement to capacity retention |